\documentclass[reprint,twocolumn,aps,prb,showpacs]{revtex4-1}

\usepackage{amsmath}
\usepackage{amssymb}
\usepackage{graphicx}
\usepackage{dcolumn}
\usepackage{bm}
\usepackage[colorlinks=true, allcolors=blue]{hyperref}
\usepackage{epstopdf}
\usepackage[utf8]{inputenc}
\usepackage{amssymb}

\begin{document}

\title{Microwave conductivity due to impurity scattering in cuprate superconductors}

\author{Minghuan Zeng, Xiang Li, Yongjun Wang, and Shiping Feng}
\email{spfeng@bnu.edu.cn}

\affiliation{Department of Physics, Beijing Normal University, Beijing 100875, China}


\begin{abstract}
The microwave surface impedance measurements on cuprate superconductors provide the crucial
information of the effect of the impurity scattering on the quasiparticle transport, however,
the full understanding of the effect of the impurity scattering on the quasiparticle transport
is still a challenging issue. Here based on the microscopic octet scattering model, the effect
of the impurity scattering on the low-temperature microwave conductivity in cuprate
superconductors is investigated in the self-consistent $T$-matrix approach. The impurity-dressed
electron propagator obtained in the Fermi-arc-tip approximation of the quasiparticle excitations
and scattering processes is employed to derive the electron current-current correlation function
by taking into account the impurity-induced vertex correction. It is shown that the microwave
conductivity spectrum is a non-Drude-like, with a sharp cusp-like peak extending to zero-energy
and a high-energy tail falling slowly with energy. Moreover, the microwave conductivity
decreases with the increase of the impurity concentration or with the increase of the strength
of the impurity scattering potential. In a striking contrast to the dome-like shape of the
doping dependence of the superconducting transition temperature, the microwave conductivity
exhibits a reverse dome-like shape of the doping dependence. The theory also show that the
highly unconventional features of the microwave conductivity are generated by both the strong
electron correlation and impurity-scattering effects.
\end{abstract}

\pacs{74.25.Nf, 74.62.Dh, 74.25.Fy, 74.72.-h}

\maketitle


\section{Introduction}

For a conventional superconductor with a s-wave pairing symmetry, the impurity scattering has
little effect on superconductivity\cite{Schrieffer64,Anderson58}. However, cuprate
superconductors are anomalously sensitive to the impurity scattering
\cite{Basov01,Hussey02,Balatsky06,Alloul09}, since superconductivity involves a paring state
with the dominant d-wave symmetry\cite{Tsuei00}. In particular, the superconducting (SC)
transition temperature $T_{\rm c}$ in cuprate superconductors is systematically diminished
with impurities
\cite{Ishida91,Legris93,Giapintzakis94,Fukuzumi96,Tolpygo96,Attfield98,Bobroff99,Eisaki04},
which therefore confirms definitely that the impurity scattering has high impacts on
superconductivity\cite{Basov01,Hussey02,Balatsky06,Alloul09}. In this case, the understanding
of the effect of the impurity scattering on superconductivity is a central issue for cuprate
superconductors.

Among the striking features of the SC-state in cuprate superconductivity, the physical quantity
which most evidently displays the dramatic effect of the impurity scattering on
superconductivity is the quasiparticle transport\cite{Basov01,Hussey02,Balatsky06,Alloul09},
which is manifested by the microwave conductivity. This microwave conductivity contains a wealth
of the information on the SC-state quasiparticle response, and is closely associated with the
superfluid density\cite{Basov01,Hussey02,Balatsky06,Alloul09}. By virtue of systematic studies
using the microwave surface impedance measurements, the low-temperature features of the
SC-state quasiparticle transport in cuprate superconductors have been well established \cite{Basov01,Hussey02,Balatsky06,Alloul09,Bonn93,Lee96,Hosseini99,Turner03,Harris06},
where an agreement has emerged that the microwave conductivity are dominated mainly by
thermally excited quasiparticles being scattered by impurities. In particular, as an evidence
of the very long-live quasiparticle excitation deep in the SC-state, the low-temperature
microwave conductivity spectrum has a cusp-like shape of the energy dependence
\cite{Bonn93,Lee96,Hosseini99,Turner03,Harris06}. However, it is still unclear how this
microwave conductivity evolves with the impurity concentration. Moreover, the experimental
observations have also shown that even minor concentrations of impurities lead to changes in
the temperature dependence of the magnetic-field penetration-depth from linear in the pure
systems to quadratic\cite{Bonn94}, while the ratio of the low-temperature superfluid density
and effective mass of the electrons $n_{\rm s}(T\rightarrow 0)/m^{*}$ is decreased when one
increases the impurity concentration\cite{Bucci94,Bernhard96,Bobroff05}.

In the d-wave SC-state of cuprate superconductors, the SC gap vanishes along the nodal
direction of the electron Fermi surface (EFS)\cite{Tsuei00}, and then as a natural
consequence, the most properties well below $T_{\rm c}$ ought to be controlled by the
quasiparticle excitations at around the nodal region of EFS. In this case, the d-wave
Bardeen-Cooper-Schrieffer (BCS) type formalism\cite{Basov01,Hussey02,Balatsky06,Alloul09},
incorporating the effect of the impurity scattering within the self-consistent $T$-matrix
approach, has been employed to study the effect of the impurity scattering on the microwave
conductivity of cuprate superconductors
\cite{Hirschfeld94,Durst00,Berlinsky00,Hettler00,Durst02,Kim04,Nunner05,Wang08}, where the
impurity scattering self-energy was evaluated in the nodal approximation of the quasiparticle
excitations and scattering processes, and then was used to calculate the electron
current-current correlation function by including the contributions of the impurity-induced
vertex correction and Fermi-liquid correction
\cite{Durst00,Berlinsky00,Hettler00,Durst02,Kim04,Nunner05}. The obtained results show that
both the impurity-induced vertex correction and Fermi-liquid correction modify the microwave
conductivity\cite{Durst00,Berlinsky00,Hettler00,Durst02,Kim04,Nunner05}.
However, (i) although the contribution from the Fermi-liquid correction is included, these
treatments suffer from ignoring the strong electron correlation effect in the homogenous part
of the electron propagator
\cite{Hirschfeld94,Durst00,Berlinsky00,Hettler00,Durst02,Kim04,Nunner05}, while this strong
electron correlation effect also plays an important role in the SC-state quasiparticle
transport; (ii) moreover, the angle-resolved photoemission spectroscopy (ARPES) experiments
\cite{Chatterjee06,He14,Restrepo23} have shown clearly that the Fermi arcs that emerge due
to the EFS reconstruction at the case of zero energy
\cite{Norman98,Shi08,Sassa11,Fujita14,Comin14,Kaminski15,Loret17,Chen19} can persist into the
case for a finite binding-energy, where a particularly large fraction of the spectral weight
is located at around the tips of the Fermi arcs. These tips of the Fermi arcs connected by the
scattering wave vectors ${\bf q}_{i}$ thus construct an {\it octet scattering model}, and then
the quasiparticle scattering with the scattering wave vectors ${\bf q}_{i}$ contribute
effectively to the quasiparticle scattering processes\cite{Chatterjee06,He14,Restrepo23}. In
particular, this octet scattering model has been employed to give a consistent explanation of
the experimental data detected from Fourier transform scanning tunneling spectroscopy
\cite{Yin21,Pan01,Kohsaka07,Kohsaka08,Hamidian16} and the ARPES autocorrelation pattern
observed from ARPES experiments\cite{Chatterjee06,He14,Restrepo23}. These experimental results
\cite{Chatterjee06,He14,Restrepo23,Norman98,Shi08,Sassa11,Fujita14,Comin14,Kaminski15,Loret17,Chen19,Yin21,Pan01,Kohsaka07,Kohsaka08,Hamidian16}
therefore have shown clearly that the shape of EFS has deep consequences for the various
properties of cuprate superconductors, while such an aspect should be also reflected in the
SC-state quasiparticle transport.

In the recent work\cite{Zeng22}, we have started from the homogenous part of the electron
propagator and the related {\it microscopic octet scattering model}, which are obtained within
the framework of the kinetic-energy-driven superconductivity
\cite{Feng0306,Feng12,Feng15,Feng15a}, to discuss the influence of the impurity scattering on
the electronic structure of cuprate superconductors in the self-consistent $T$-matrix approach,
where the impurity scattering self-energy is derived in the {\it Fermi-arc-tip approximation}
of the quasiparticle excitations and scattering processes, and then the impurity-dressed
electron propagator incorporates both the strong electron correlation effect and the
impurity-scattering effect. The obtained results\cite{Zeng22} show that the decisive role
played by the impurity scattering self-energy in the particle-hole channel is the further
renormalization of the quasiparticle band structure with a reduced quasiparticle lifetime,
while the impurity scattering self-energy in the particle-particle channel induces a strong
deviation from the d-wave behaviour of the SC gap, leading to the existence of a finite gap
over the entire EFS. In this paper, we study the effect of the impurity scattering on the
microwave conductivity in cuprate superconductors along with this line by taking into
account the impurity-induced vertex correction, where the impurity-dressed electron
propagator\cite{Zeng22} is employed to evaluate the vertex-corrected electron
current-current correlation function in the self-consistent $T$-matrix approach, and the
obtained results in the Fermi-arc-tip approximation of the quasiparticle excitations and
scattering processes show that the low-temperature microwave conductivity spectrum is a
non-Drude-like, with a sharp cusp-like peak extending to zero-energy and a high-energy tail
falling slowly with energy, in agreement with the corresponding experiments
\cite{Bonn93,Lee96,Hosseini99,Turner03,Harris06}. In particular, although the low-energy
cusp-like peak decay as $\rightarrow 1/[\omega+{\rm constant}]$, the overall shape of the
microwave conductivity spectrum exhibits a special non-Drude-like behavior with the depicted
formula that has been also used to fit the corresponding experimental data in
Ref. \onlinecite{Turner03}. Moreover, the microwave conductivity decreases with ascending
impurity concentration or with rising strength of the impurity scattering potential. Our
these results therefore show that the highly unconventional features of the microwave
conductivity are induced by both the strong electron correlation and impurity-scattering
effects.

The remainder of this paper is organized as follows: Sec. \ref{Formalism} contains details
regarding the calculation technique of the microwave conductivity in the presence of the
impurity scattering. The quantitative characteristics of the impurity-scattering effect on
the doping and energy dependence of the microwave conductivity are presented in
Sec. \ref{Quantitative-characteristics}, where we show that in a striking contrast to the
dome-like shape doping dependence of $T_{\rm c}$, the minimum of the microwave conductivity
occurs at around the optimal doping, and then increases in both underdoped and overdoped
regimes. Finally, we give a summary in Sec. \ref{conclude}. In the Appendix, we present the
details of the derivation of the vertex kernels of the electron current-current correlation
function.

\section{Theoretical Framework}\label{Formalism}

It was recognized shortly after the discovery of superconductivity in cuprate superconductors
that the essential physics of cuprate superconductors is contained in the square-lattice
$t$-$J$ model\cite{Anderson87,Zhang88},
\begin{eqnarray}\label{tjham}
H&=&-\sum_{ll'\sigma}t_{ll'}C^{\dagger}_{l\sigma}C_{l'\sigma}
+\mu\sum_{l\sigma}C^{\dagger}_{l\sigma}C_{l\sigma}
+J\sum_{l\hat{\eta}}{\bf S}_{l}\cdot {\bf S}_{l+\hat{\eta}},~~~~
\end{eqnarray}
where $C^{\dagger}_{l\sigma}$ ($C_{l\sigma}$) creates (annihilates) a constrained electron
with spin index $\sigma=\uparrow,\downarrow$ on lattice site $l$, ${\bf S}_{l}$ is spin operator
with its components $S^{\rm x}_{l}$, $S^{\rm y}_{l}$, and $S^{\rm z}_{l}$, and $\mu$ is the
chemical potential. The kinetic-energy part includes the electron-hopping term
$t_{ll'}=t_{\hat{\eta}}=t$ between the nearest-neighbor (NN) sites $\hat{\eta}$ and the
electron-hopping term $t_{ll'}=t_{\hat{\eta}'}=t'$ between the next NN sites $\hat{\eta}'$,
while the magnetic-energy part is described by a Heisenberg term with the magnetic interaction
$J$ between the NN sites $\hat{\eta}$. As a qualitative discussion, the commonly used parameters
in the $t$-$J$ model (\ref{tjham}) are chosen as $t/J=2.5$ and $t'/t=0.3$ as in our previous
discussions \cite{Zeng22}. However, when necessary to compare with the experimental data, we set
$J=1000$K.

The basis set of the $t$-$J$ model (\ref{tjham}) is restricted by the requirement that no
lattice site may be doubly occupied by electrons\cite{Yu92,Feng93,Zhang93,Guillou95}, i.e.,
$\sum_{\sigma}C^{\dagger}_{l\sigma}C_{l\sigma}\leq 1$. Our method employs a fermion-spin theory
description of the $t$-$J$ model (\ref{tjham}) together with the on-site local constraint of no
double electron occupancy \cite{Feng15,Feng0494}, where the constrained electron operators
$C_{l\uparrow}$ and $C_{l\downarrow}$ in the $t$-$J$ model (\ref{tjham}) are separated into two
distinct operators as,
\begin{eqnarray}\label{CSSFS}
C_{l\uparrow}=h^{\dagger}_{l\uparrow}S^{-}_{l},~~~~
C_{l\downarrow}=h^{\dagger}_{l\downarrow}S^{+}_{l},
\end{eqnarray}
with the spinful fermion operator $h_{l\sigma}=e^{-i\Phi_{l\sigma}}h_{l}$ that describes the
charge degree of freedom of the constrained electron together with some effects of spin
configuration rearrangements due to the presence of the doped hole itself (charge carrier),
while the spin operator $S_{l}$ that represents the spin degree of freedom of the constrained
electron, and then the local constraint of no double electron occupancy is fulfilled in actual
analyses.

Starting from the $t$-$J$ model (\ref{tjham}) in the fermion-spin representation (\ref{CSSFS}),
the kinetic-energy-driven SC mechanism has been established
\cite{Feng15,Feng0306,Feng12,Feng15a}, where the charge carriers are held together in the
d-wave pairs in the particle-particle channel due to the effective interaction, which
originates directly from the kinetic energy of the $t$-$J$ model (\ref{tjham}) in the
fermion-spin representation (\ref{CSSFS}) by the exchange of spin excitations, then the d-wave
electron pairs originating from the d-wave charge-carrier pairing state are due to the
charge-spin recombination, and their condensation reveals the d-wave SC-state. In these
previous discussions, the homogenous electron propagator of the $t$-$J$ model (\ref{tjham}) in
the SC-state has been obtained explicitly in the Nambu representation as\cite{Feng15a},
\begin{eqnarray}\label{EGF-NR}
\tilde{G}({\bf k},\omega)&=&\left(
\begin{array}{cc}
G({\bf k},\omega), & \Im({\bf k},\omega) \\
\Im^{\dagger}({\bf k},\omega), & -G({\bf k},-\omega)
\end{array}\right)\nonumber\\
&=&{1\over F({\bf k},\omega)}\{[\omega-\Sigma_{0}({\bf k},\omega)]\tau_{0}
+\Sigma_{1}({\bf k},\omega)\tau_{1}\nonumber\\
&+&\Sigma_{2}({\bf k},\omega)\tau_{2}+[\varepsilon_{\bf k}
+\Sigma_{3}({\bf k},\omega)]\tau_{3}\},~~~
\end{eqnarray}
where $\tau_{0}$ is the unit matrix, $\tau_{1}$, $\tau_{2}$, and $\tau_{3}$ are Pauli
matrices, $\varepsilon_{\bf k}=-4t\gamma_{\bf k}+4t'\gamma_{\bf k}'+\mu$ is the energy
dispersion in the tight-binding approximation, with
$\gamma_{\bf k}=({\rm cos}k_{x}+{\rm cos} k_{y})/2$,
$\gamma_{\bf k}'={\rm cos}k_{x}{\rm cos}k_{y}$,
$F({\bf k},\omega)=[\omega-\Sigma_{0}({\bf k},\omega)]^{2}-[\varepsilon_{\bf k}
+\Sigma_{3}({\bf k},\omega)]^{2}-\Sigma^{2}_{1}({\bf k},\omega)
-\Sigma^{2}_{2}({\bf k},\omega)$, and the homogenous self-energy has been expanded into
its constituent Pauli matrix components as,
\begin{eqnarray}\label{ESE-NR}
&~&\tilde{\Sigma}({\bf k},\omega)=\sum_{\alpha=0}^{3}
\Sigma_{\alpha}({\bf k},\omega)\tau_{\alpha}\nonumber\\
&=&\left(
\begin{array}{cc}
\Sigma_{0}({\bf k},\omega)+\Sigma_{3}({\bf k},\omega),
&\Sigma_{1}({\bf k},\omega)-i\Sigma_{2}({\bf k},\omega)\\
\Sigma_{1}({\bf k},\omega)+i\Sigma_{2}({\bf k},\omega),
&\Sigma_{0}({\bf k},\omega)-\Sigma_{3}({\bf k},\omega)
\end{array}\right),~~~~~
\end{eqnarray}
with $\Sigma_{0}({\bf k},\omega)$ and $\Sigma_{3}({\bf k},\omega)$ that are respectively the
antisymmetric and symmetric parts of the homogenous self-energy in the particle-hole channel,
while $\Sigma_{1}({\bf k},\omega)$ and $\Sigma_{2}({\bf k},\omega)$ that are respectively the
real and imaginary parts of the homogenous self-energy in the particle-particle channel.
Moreover, these homogenous self-energies $\Sigma_{0}({\bf k},\omega)$,
$\Sigma_{1}({\bf k},\omega)$, $\Sigma_{2}({\bf k},\omega)$, and $\Sigma_{3}({\bf k},\omega)$
have been derived explicitly in Ref. \onlinecite{Feng15a} in terms of the full charge-spin
recombination. In particular, the sharp peaks visible for temperature $T\rightarrow 0$ in
$\Sigma_{0}({\bf k},\omega)$, $\Sigma_{1}({\bf k},\omega)$, $\Sigma_{2}({\bf k},\omega)$, and
$\Sigma_{3}({\bf k},\omega)$ are actually a $\delta$-functions, broadened by a small damping
used in the numerical calculation for a finite lattice. The calculation in this paper for
$\Sigma_{0}({\bf k},\omega)$, $\Sigma_{1}({\bf k},\omega)$, $\Sigma_{2}({\bf k},\omega)$, and
$\Sigma_{3}({\bf k},\omega)$ is performed numerically on a $120\times 120$ lattice in momentum
space, with the infinitesimal $i0_{+}\rightarrow i\Gamma$ replaced by a small damping
$\Gamma=0.05J$.

The homogenous electron spectral function can be obtained directly from the above homogenous
electron propagator (\ref{EGF-NR}). In this case, the topology of EFS in the pure system has
been discussed in terms of the intensity map of the homogenous electron spectral function at
zero energy\cite{Liu21,Gao18,Gao19}, and the obtained results show that EFS contour is broken
up into the disconnected Fermi arcs located around the nodal
region\cite{Norman98,Shi08,Sassa11,Fujita14,Comin14,Kaminski15,Loret17,Chen19}, however, a
large number of the low-energy electronic states is available at around the tips of the Fermi
arcs, and then all the anomalous properties arise from these quasiparticle excitations
located at around the tips of the Fermi arcs. In particular, these tips of the Fermi arcs
connected by the scattering wave vectors ${\bf q}_{i}$ naturally construct an {\it octet
scattering model}, and then the quasiparticle scattering with the scattering wave vectors
${\bf q}_{i}$ therefore contribute effectively to the quasiparticle scattering processes
\cite{Yin21,Pan01,Kohsaka07,Kohsaka08,Hamidian16}. Moreover, this {\it octet scattering model}
can persist into the case for a finite binding-energy \cite{Chatterjee06,He14,Restrepo23},
which leads to that the sharp peaks in the ARPES autocorrelation spectrum with the scattering
wave vectors ${\bf q}_{i}$ are directly correlated to the regions of the highest joint density
of states.

\subsection{Impurity-dressed electron propagator}\label{dressed-propagator}

In the low-temperature limit, the framework for the discussions of the impurity-scattering
effect is the self-consistent $T$-matrix approach
\cite{Basov01,Hussey02,Balatsky06,Alloul09,Mahan81,Hirschfeld89,Hirschfeld93}. The discussions
of the low-temperature microwave conductivity of cuprate superconductors in this paper builds
on the impurity-dressed electron propagator, which is obtained from the dress of the homogenous
electron propagator (\ref{EGF-NR}) via the impurity scattering\cite{Zeng22}, where the
self-consistent $T$-matrix approach is employed to derive the impurity scattering self-energy
in the Fermi-arc-tip approximation of the quasiparticle excitations and scattering processes.
For a convenience in the following discussions of the microwave conductivity, a short summary
of the derivation process of the impurity-dressed electron propagator \cite{Zeng22} is
therefore given in this subsection.

The homogenous electron propagator in Eq. (\ref{EGF-NR}) is dressed due to the presence of the
impurity scattering\cite{Basov01,Hussey02,Balatsky06,Alloul09}, and can be expressed explicitly
as,
\begin{eqnarray}\label{ID-EGF-1}
\tilde{G}_{\rm I}({\bf k},\omega)^{-1}=\tilde{G}({\bf k},\omega)^{-1}
-\tilde{\Sigma}_{\rm I}({\bf k},\omega),
\end{eqnarray}
where in a striking similarity to the homogenous self-energy (\ref{ESE-NR}), the impurity
scattering self-energy $\tilde{\Sigma}_{\rm I}({\bf k},\omega)$ can be also expanded into its
constituent Pauli matrix components as,
\begin{eqnarray}\label{ESE-IS}
&~&\tilde{\Sigma}_{\rm I}({\bf k},\omega)=\sum_{\alpha=0}^{3}
\Sigma_{{\rm I}\alpha}({\bf k},\omega)\tau_{\alpha}\nonumber\\
&=&\left(
\begin{array}{cc}
\Sigma_{\rm I0}({\bf k},\omega)+\Sigma_{\rm I3}({\bf k},\omega),
&\Sigma_{\rm I1}({\bf k},\omega)-i\Sigma_{\rm I2}({\bf k},\omega)\\
\Sigma_{\rm I1}({\bf k},\omega)+i\Sigma_{\rm I2}({\bf k},\omega),
&\Sigma_{\rm I0}({\bf k},\omega)-\Sigma_{\rm I3}({\bf k},\omega)
\end{array}\right).~~~~~
\end{eqnarray}
The above impurity scattering self-energy together with the dressed electron propagator
(\ref{ID-EGF-1}) can be analyzed in the self-consistent $T$-matrix approach
\cite{Mahan81,Hirschfeld89,Hirschfeld93}, where $\tilde{\Sigma}_{\rm I}({\bf k},\omega)$ can
be derived approximately as,
\begin{eqnarray}\label{SE-FIS-1}
\tilde{\Sigma}_{\rm I}({\bf k},\omega)=n_{\rm i}N\tilde{T}_{{\bf k}{\bf k}}(\omega),
\end{eqnarray}
with the impurity concentration $n_{\rm i}$, the number of sites on a square lattice $N$, and
the diagonal part of the T-matrix $\tilde{T}_{{\bf k}{\bf k}}(\omega)$, while the
self-consistent T-matrix equation that can be expressed formally by the summation of all
impurity scattering processes as,
\begin{equation}\label{TMAT-ORI}
\tilde{T}_{{\bf k}{\bf k}'}={1\over N}\tau_{3}V_{{\bf k}{\bf k}'}+{1\over N}\sum_{{\bf k}''}
V_{{\bf k}{\bf k}''}\tau_{3}\tilde{G}_{\rm I}({\bf k}'',\omega)\tilde{T}_{{\bf k}''{\bf k}'},
\end{equation}
where $V_{{\bf k}{\bf k}'}$ is the momentum dependence of the impurity scattering potential. It
thus shows that the initial and final momenta of an impurity scattering event must always be
equal to the momentum-space sited in the Brillouin zone (BZ).

\begin{figure}[h!]
\centering
\includegraphics[scale=0.3]{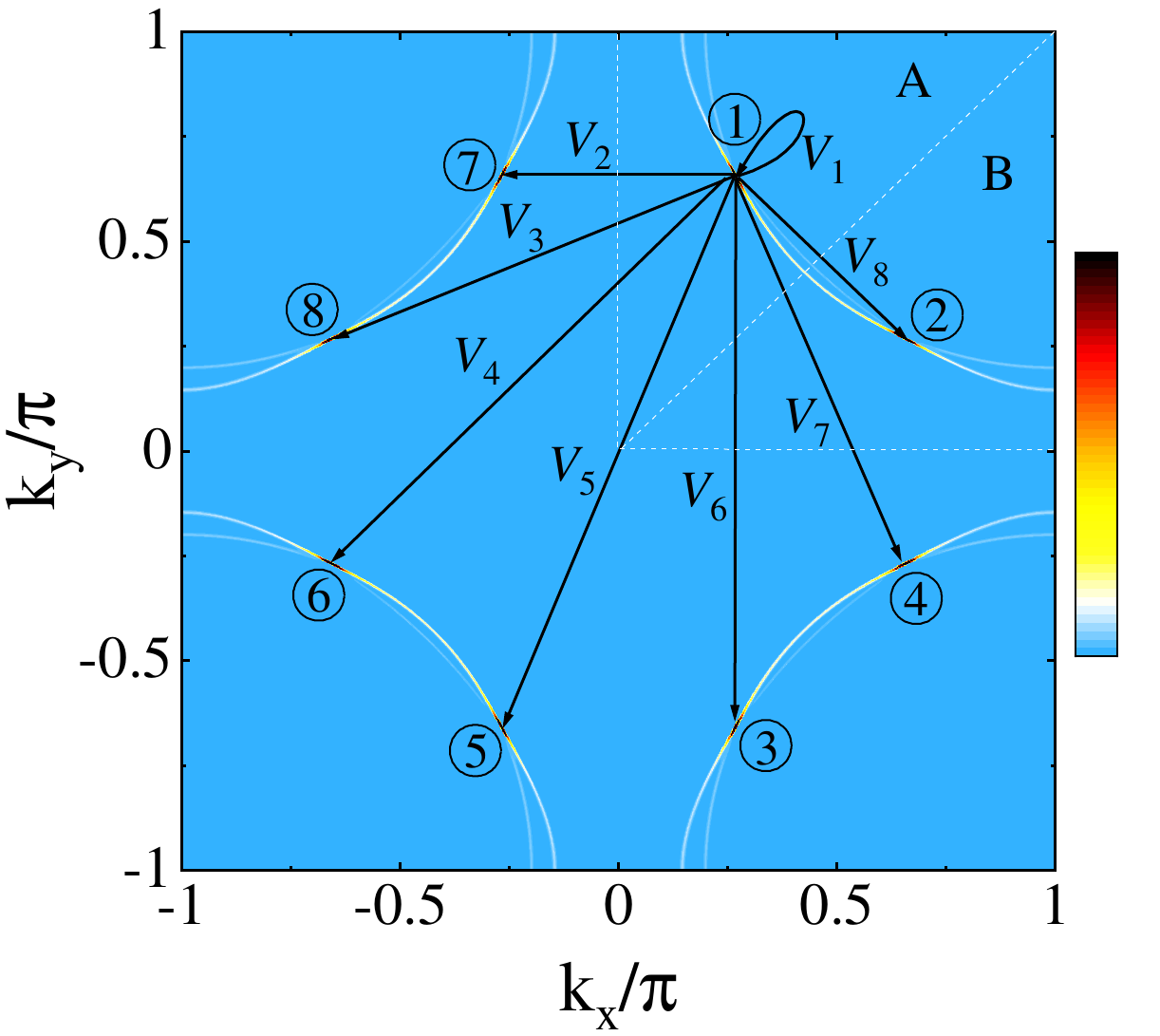}
\caption{(Color online) The impurity scattering in the microscopic octet scattering model,
where $V_{1}$ is the impurity scattering potential for the intra-tip scattering, $V_{2}$,
$V_{3}$, $V_{7}$, and $V_{8}$ are the impurity scattering potentials for the adjacent-tip
scattering, while $V_{4}$, $V_{5}$, and $V_{6}$ are the impurity scattering potentials for
the opposite-tip scattering. In the d-wave superconducting-state, the tips of the Fermi arcs
are divided into two groups: (A) the tips of the Fermi arcs located at the region of
$|k_{y}|>|k_{x}|$ and (B) the tips of the Fermi arcs located at the region of
$|k_{x}|>|k_{y}|$. \label{Fermi-arc-tip-picture}}
\end{figure}

However, in the microscopic octet scattering model\cite{Zeng22} shown in
Fig. \ref{Fermi-arc-tip-picture}, a particularly large fraction of the spectral weight is
accommodated at around eight tips of the Fermi arcs in the case of low temperatures and low
energies, indicating that a large number of the quasiparticle excitations are induced only at
around these eight tips of the Fermi arcs. On the other hand, the strength of the impurity
scattering potential $V_{{\bf k}{\bf k}'}$ in the T-matrix equation (\ref{TMAT-ORI}) falls
off quickly when the momentum shifts away from the tips of the Fermi arcs. In this case, the
initial and final momenta of an impurity scattering event are always approximately equal to the
momentum-space sited at around one of these eight tips of the Fermi arcs. In this Fermi-arc-tip
approximation\cite{Zeng22}, we only need to consider three possible cases as shown in
Fig. \ref{Fermi-arc-tip-picture} for the impurity scattering potential $V_{{\bf k}{\bf k}'}$
in the $T$-matrix equation (\ref{TMAT-ORI}): (i) the impurity scattering potential for the
scattering process at the intra-tip of the Fermi arc $V_{{\bf k}{\bf k}'}=V_{1}$, where
${\bf k}$ and ${\bf k}'$ are located at the same tip of the Fermi arc; (ii) the impurity
scattering potentials for the scattering process at the adjacent-tips of the Fermi arcs
$V_{{\bf k}{\bf k}'}=V_{2}$, $V_{{\bf k}{\bf k}'}=V_{3}$, $V_{{\bf k}{\bf k}'}=V_{7}$, and
$V_{{\bf k}{\bf k}'}=V_{8}$, where ${\bf k}$ and ${\bf k}'$ are located at the adjacent-tips of
the Fermi arcs; (iii) the impurity scattering potentials for the scattering process at the
opposite-tips of the Fermi arcs $V_{{\bf k}{\bf k}'}=V_{4}$, $V_{{\bf k}{\bf k}'}=V_{5}$, and
$V_{{\bf k}{\bf k}'}=V_{6}$, where ${\bf k}$ and ${\bf k}'$ are located at the opposite-tips
of the Fermi arcs, and then the impurity scattering potential $V_{{\bf k}{\bf k}'}$ in the
self-consistent T-matrix equation (\ref{TMAT-ORI}) is reduced as a $8\times 8$-matrix,
\begin{eqnarray}\label{ISP-matrix}
\tilde{V} =\left(
\begin{array}{cccc}
V_{11} & V_{12} & \cdots & V_{18}\\
V_{21} & V_{22} & \cdots & V_{28}\\
\vdots & \vdots & \ddots & \vdots\\
V_{81} & V_{82} & \cdots & V_{88}
\end{array}\right),
\end{eqnarray}
where the matrix elements are given by: $V_{jj}=V_{1}$ for $j=1,2,3,... 8$,
$V_{jj'}=V_{j'j}=V_{2}$ for $j=1,2,3,6$ with the corresponding $j'=7,4,5,8$, respectively,
$V_{jj'}=V_{j'j}=V_{3}$ for $j=1,2,3,4$ with the corresponding $j'=8,7,6,5$, respectively,
$V_{jj'}=V_{jj'}=V_{4}$ for $j=1,2,3,4$ with the corresponding $j'=6,5,8,7$, respectively,
$V_{jj'}=V_{j'j}=V_{5}$ for $j=1,2,3,4$ with the corresponding $j'=5,6,7,8$, respectively,
$V_{jj'}=V_{j'j}=V_{6}$ for $j=1,2,4,5$ with the corresponding $j'=3,8,6,7$, respectively,
$V_{jj'}=V_{j'j}=V_{7}$ for $j=1,2,5,6$ with the corresponding $j'=4,3,8,7$, respectively,
and $V_{jj'}=V_{j'j}=V_{8}$, for $j=1,3,5,7$ with the corresponding $j'=2,4,6,8$,
respectively.

With the help of the above impurity scattering potential matrix $\tilde{V}$, the
self-consistent T-matrix equation (\ref{TMAT-ORI}) is reduced as a $16\times 16$-matrix
equation around eight tips of the Fermi arcs as,
\begin{equation}\label{T-matrix-ISP}
\tilde{T}_{jj'}={1\over N}\tau_{3}V_{jj'}+{1\over N}\sum_{j''{\bf k}''}V_{jj''}[\tau_{3}
\tilde{G}_{\rm I}({\bf k}'',\omega)]\tilde{T}_{j''j'},
\end{equation}
where $j$, $j'$, and $j''$ label the tips of the Fermi arcs, the summation ${\bf k}''$ is
restricted within the area around the tip $j''$ of the Fermi arc, $\tilde{T}_{jj'}$ is now an
impurity-average quantity, and then the impurity
scattering self-energy $\tilde{\Sigma}_{\rm I}({\bf k},\omega)$ in Eq. (\ref{SE-FIS-1}) is
obtained as,
\begin{eqnarray}\label{SE-FIS}
\tilde{\Sigma}_{\rm I}(\omega)=n_{\rm i}N\tilde{T}_{jj}(\omega).
\end{eqnarray}

It has been shown that the diagonal propagator in Eq. (\ref{ID-EGF-1}) is symmetrical about the
nodal direction, while the off-diagonal propagator is asymmetrical about the nodal direction,
since the SC-state has a d-wave symmetry\cite{Zeng22}. In this case, the region of the location
of the tips of the Fermi arcs has been separated into two groups: (A) the tips of the Fermi
arcs located at the region of $|k_{y}|>|k_{x}|$, and (B) the tips of the Fermi arcs located at
the region of $|k_{x}|>|k_{y}|$, and then the dressed electron propagator
$\tilde{G}_{I}({\bf k},\omega)$ in Eq. (\ref{ID-EGF-1}) can be also derived in the regions A
and B as\cite{Zeng22},
\begin{subequations}\label{ID-EGF-AB}
\begin{eqnarray}
&&\tilde{G}^{\rm (A)}_{\rm I}({\bf k},\omega)
={1\over F^{\rm (A)}_{\rm I}({\bf k},\omega)}\{[\omega-\Sigma_{0}({\bf k},\omega)
-\Sigma_{\rm I0}(\omega)]\tau_{0}\nonumber\\
&&+[\Sigma_{1}({\bf k},\omega)+\Sigma^{\rm (A)}_{\rm I1}(\omega)]\tau_{1}
+[\Sigma_{2}({\bf k},\omega)+\Sigma^{\rm (A)}_{\rm I2}(\omega)]\tau_{2}\nonumber\\
&&+[\varepsilon_{\bf k}+\Sigma_{3}({\bf k},\omega)+\Sigma_{\rm I3}(\omega)]
\tau_{3}\},~~\\
&&\tilde{G}^{\rm (B)}_{\rm I}({\bf k},\omega)
={1\over F^{\rm (B)}_{\rm I}({\bf k},\omega)}\{[\omega-\Sigma_{0}({\bf k},\omega)
-\Sigma_{\rm I0}(\omega)]\tau_{0}\nonumber\\
&&+[\Sigma_{1}({\bf k},\omega)+\Sigma^{\rm (B)}_{\rm I1}(\omega)]\tau_{1}
+[\Sigma_{2}({\bf k},\omega)+\Sigma^{\rm (B)}_{\rm I2}(\omega)]\tau_{2}\nonumber\\
&&+[\varepsilon_{\bf k}+\Sigma_{3}({\bf k},\omega)+\Sigma_{\rm I3}(\omega)]
\tau_{3}\},~~
\end{eqnarray}
\end{subequations}
respectively, where $F^{\rm (A)}_{\rm I}({\bf k},\omega)=[\omega-\Sigma_{0}({\bf k},\omega)
-\Sigma_{\rm I0}(\omega)]^{2}-[\varepsilon_{\bf k}+\Sigma_{3}({\bf k},\omega)
+\Sigma_{\rm I3}(\omega)]^{2}-[\Sigma_{1}({\bf k},\omega)
+\Sigma^{\rm (A)}_{\rm I1}(\omega)]^{2}-[\Sigma_{2}({\bf k},\omega)
+\Sigma^{\rm (A)}_{\rm I2}(\omega)]^{2}$, $F^{\rm (B)}_{\rm I}({\bf k},\omega)
=[\omega-\Sigma_{0}({\bf k},\omega)
-\Sigma_{\rm I0}(\omega)]^{2}-[\varepsilon_{\bf k}+\Sigma_{3}({\bf k},\omega)
+\Sigma_{\rm I3}(\omega)]^{2}-[\Sigma_{1}({\bf k},\omega)
+\Sigma^{\rm (B)}_{\rm I1}(\omega)]^{2}-[\Sigma_{2}({\bf k},\omega)
+\Sigma^{\rm (B)}_{\rm I2}(\omega)]^{2}$. In the self-consistent $T$-matrix approach, these
impurity scattering self-energies
$\Sigma^{\rm (A)}_{\rm I0}(\omega)$ [$\Sigma^{\rm (B)}_{\rm I0}(\omega)$],
$\Sigma^{\rm (A)}_{\rm I1}(\omega)$ [$\Sigma^{\rm (B)}_{\rm I1}(\omega)$],
$\Sigma^{\rm (A)}_{\rm I2}(\omega)$ [$\Sigma^{\rm (B)}_{\rm I2}(\omega)$], and
$\Sigma^{\rm (A)}_{\rm I3}(\omega)$ [$\Sigma^{\rm (B)}_{\rm I3}(\omega)$] and the related
$T$-matrix $\tilde{T}^{\rm (A)}_{jj'}=\sum_{\alpha}T^{(\alpha)}_{{\rm A}jj'}\tau_{\alpha}$
[$\tilde{T}^{\rm (B)}_{jj'}=\sum_{\alpha}T^{(\alpha)}_{{\rm B}jj'}\tau_{\alpha}$] with the
matrix elements $T^{(\alpha)}_{{\rm A}jj'}$ [$T^{(\alpha)}_{{\rm B}jj'}$] in
Eq. (\ref{T-matrix-ISP}) have been obtained in the Fermi-arc-tip approximation of the
quasiparticle excitations and scattering processes, and given explicitly in
Ref. \onlinecite{Zeng22}.

With the help of the above dressed electron propagator (\ref{ID-EGF-AB}) [then the dressed
electron spectral function], we\cite{Zeng22} have also discussed the influence of the
impurity scattering on the electronic structure of cuprate superconductors, and the
obtained results of the line-shape in the quasiparticle excitation spectrum and the ARPES
autocorrelation spectrum are well consistent with the corresponding experimental results
\cite{Dessau91,Hwu91,Randeria95,Fedorov99,Lu01,Sakai13,DMou17,Chatterjee06,He14,Restrepo23}.

\subsection{Microwave Conductivity}\label{MConductivity}

Now we turn to derive the microscopic conductivity of cuprate superconductors in the presence
of impurities, which is closely associated with the dressed electron propagator
(\ref{ID-EGF-AB}). The linear response theory allows one to obtain the microwave conductivity
in terms of the Kubo formula\cite{Mahan81},
\begin{equation}\label{conductivity-1}
\mathop{\sigma}\limits^{\leftrightarrow}(\Omega,T) =
-{{\rm Im}{\mathop{\Pi}\limits^{\leftrightarrow}}(\Omega)\over\Omega},
\end{equation}
where ${\mathop{\Pi}\limits^{\leftrightarrow}}(\Omega)$ is the retarded electron
current-current correlation function, and can be expressed explicitly as,
\begin{equation}\label{corP-1}
\mathop{\Pi}^{\leftrightarrow}(i\Omega_{m})=-{1\over N}\int_{0}^{\beta}d\tau e^{i\Omega_{m}\tau}
\langle T_{\tau}\bm{J}(\tau)\bm{J}(0)\rangle,
\end{equation}
with $\beta = 1/T$, the bosonic Matsubara frequency $\Omega_{m}=2\pi m/\beta$, and the current
density of electrons $\bm{J}$. This current density of electrons can be obtained in terms of
the electron polarization operator, which is a summation over all the particles and their
positions\cite{Mahan81},
and can be expressed explicitly in the fermion-spin representation (\ref{CSSFS}) as
${\bf P}=\sum_{l\sigma}{\bf R}_{l}\hat{C}^{\dag}_{l\sigma}\hat{C}_{l\sigma}={1\over 2}
\sum_{l\sigma}{\bf R}_{l}h_{l\sigma}h^{\dag}_{l\sigma}$. Within the $t$-$J$ model (\ref{tjham})
in the fermion-spin representation (\ref{CSSFS}), the current density of electrons is obtained
by evaluating the time-derivative of the polarization operator using the Heisenberg's equation
of motion as,
\begin{widetext}
\begin{eqnarray}\label{current-density-1}
{\bf J} &=& -ie[H,{\bf P}]
= -i{1\over 2}et\sum_{\langle l\hat{\eta}\rangle}\hat{\eta}(h^{\dagger}_{l+\hat{\eta}\uparrow}
h_{l\uparrow}S_{l}^{+}S^{-}_{l+\hat{\eta}}+h^{\dagger}_{l+\hat{\eta}\downarrow}h_{l\downarrow}
S^{\dagger}_{l}S^{-}_{l+\hat{\eta}})
+i{1\over 2}et'\sum_{\langle l\hat{\eta}'\rangle}\hat{\eta}'(h^{\dagger}_{l+\hat{\eta}'\uparrow}
h_{l\uparrow}S_{l}^{+}S^{-}_{l+\hat{\eta}'}+h^{\dagger}_{l+\hat{\eta}'\downarrow}h_{l\downarrow}
S^{\dagger}_{l}S^{-}_{l+\hat{\eta}'})\nonumber \\
&=&i{1\over 2}et\sum_{\langle l\hat{\eta}\rangle\sigma}\hat{\eta}C^{\dagger}_{l\sigma}
C_{l+\hat{\eta}\sigma}-i{1\over 2}et'\sum_{\langle l\hat{\eta}'\rangle\sigma}\hat{\eta}'
C^{\dagger}_{l\sigma}C_{l+\hat{\eta}'\sigma}
\approx -e\bm{V}_{\rm F}\sum\limits_{\bm{k}\sigma}C^{\dagger}_{{\bf k}\sigma}
C_{{\bf k}\sigma},
\end{eqnarray}
\end{widetext}
with the electron charge $e$, the electron Fermi velocity $\bm{V}_{\rm F}$, which can be
derived directly from the energy dispersion $\varepsilon_{\bf k}$ in the tight-binding
approximation in Eq. (\ref{EGF-NR}) as,
\begin{eqnarray}\label{electron-velocity-1}
\bm{V}_{\rm F} = V^{(x)}_{\rm F}\hat{k}_{x}+V^{(y)}_{\rm F}\hat{k}_{y}
= V_{\rm F}[\hat{k}_{x}\cos\theta_{{\bm{k}}_{\rm F}}
+\hat{k}_{y}\sin\theta_{{\bm{k}}_{\rm F}}],~~~~~~
\end{eqnarray}
where $V^{(x)}_{\rm F}=t\sin k^{(x)}_{\rm F}-2t'\sin k^{(x)}_{\rm F}\cos k^{(y)}_{\rm F}$,
$V^{(y)}_{\rm F}=t\sin k^{(y)}_{\rm F}-2t'\sin k^{(y)}_{\rm F}\cos k^{(x)}_{\rm F}$,
$\cos\theta_{\bm{k}_{\rm F}}=V^{(x)}_{\rm F}/V_{\rm F}$,
$\sin\theta_{\bm{k}_{\rm F}}=V^{(y)}_{\rm F}/V_{\rm F}$, and
$V_{\rm F}=\sqrt{[V^{(x)}_{\rm F}]^{2}+[V^{(y)}_{\rm F}]^{2}}$.
For a convenience in the following discussions of the electron current-current correlation
function (\ref{corP-1}), the electron operators can be rewritten in the Nambu representation
as $\Psi^{\dagger}_{\bf k}=(C^{\dagger}_{{\bf k}\uparrow},C_{-{\bf k}\downarrow})$ and
$\Psi_{\bf k}=(C_{{\bf k}\uparrow},C^{\dagger}_{-{\bf k}\downarrow})^{\rm T}$, and then the
current density of electrons in Eq. (\ref{current-density-1}) can be rewritten in the Nambu
representation as,
\begin{eqnarray}\label{current-density-200}
\bm{J} = -e\bm{V}_{\rm F}\sum\limits_{\bm{k}}\Psi_{\bm{k}}^{\dagger}
\tau_{0}\Psi_{\bm{k}}.
\end{eqnarray}

With the help of the above current density of electrons (\ref{current-density-200}), the
impurity-induced vertex-corrected current-current correlation function (\ref{corP-1}) can be
formally expressed in terms of the dressed electron propagator as,
\begin{widetext}
\begin{eqnarray}\label{corP-3}
{\mathop{\Pi}^{\leftrightarrow}}(i\Omega_{m})={1\over N}\int_{0}^{\beta}d\tau
e^{i\Omega_{m}\tau}{\mathop{\Pi}^{\leftrightarrow}}(\tau)
=(eV_{\rm F})^{2}{1\over N}\sum\limits_{\bm{k}}{1\over\beta}\sum\limits_{i\omega_{n}}
\hat{\bm{k}}Tr[\tilde{G}_{\text{I}}(\bm{k},i\omega_{n})
\tilde{G}_{\text{I}}(\bm{k},i\omega_{n}+i\Omega_{m})
\tilde{\Gamma}(\bm{k},i\omega_{n},i\Omega_{m})],
\end{eqnarray}
where $\omega_{n}=(2n+1)\pi/\beta$ is the fermionic Matsubara frequency, while the
impurity-induced vertex correction in the ladder approximation can be generally expressed
as\cite{Mahan81},
\begin{eqnarray}\label{T-matrix-equation-1}
\tilde{\Gamma}(\bm{k},i\omega_{n},i\Omega_{m})=\hat{\bm{k}}\tau_{0}+n_{i}N\sum_{\bm{k}''}
\tilde{T}_{\bm{k}\bm{k}''}(i\omega_{n}+i\Omega_{m})
\tilde{G}_{\text{I}}(\bm{k}'',i\omega_{n}+i\Omega_{m})
\tilde{\Gamma}(\bm{k}'',i\omega_{n},i\Omega_{m})\tilde{G}_{\text{I}}(\bm{k}'',i\omega_{n})
\tilde{T}_{\bm{k}''\bm{k}}(i\omega_{n}).
\end{eqnarray}

Starting from the homogenous part of the d-wave BCS type formalism, the effect of the impurity
scattering on the microwave conductivity has been discussed in the self-consistent $T$-matrix
approach by taking into account the impurity-induced vertex correction
\cite{Durst00,Berlinsky00,Hettler00,Durst02,Kim04,Nunner05}, where the vertex-corrected
electron current-current correlation function and the related impurity-dressed electron
propagator have been evaluated in the {\it nodal approximation}. In the following discussions,
the vertex-corrected electron current-current correlation function is generalized from the
previous case obtained in the {\it nodal approximation}
\cite{Durst00,Berlinsky00,Hettler00,Durst02,Kim04,Nunner05} to the present case in the
{\it Fermi-arc-tip approximation}, where the impurity-induced vertex correction for the
electron current-current correlation function (\ref{T-matrix-equation-1}) can be expressed
explicitly in the regions A and B as,
\begin{subequations}\label{Vertex-Fun}
\begin{eqnarray}
\tilde{\Gamma}^{(\text{A})}(\bm{k},i\omega_{n},i\Omega_{m})&=&\hat{k}_{\rm F}^{(j)}\tau_{0}
+\hat{k}_{x}^{(j)}\tilde{\Lambda}^{(\text{A})}_{x}(i\omega_{n},i\Omega_{m})+\hat{k}_{y}^{(j)}
\tilde{\Lambda}^{(\text{A})}_{y}(i\omega_{n},i\Omega_{m}),~~~~~{\rm for}~j\in {\rm odd},
\label{Vertex-Fun-A}\\
\tilde{\Gamma}^{(\text{B})}(\bm{k},i\omega_{n},i\Omega_{m})&=&\hat{k}_{\rm F}^{(j)}\tau_{0}
+\hat{k}_{x}^{(j)}\tilde{\Lambda}^{(\text{B})}_{x}(i\omega_{n},i\Omega_{m})+\hat{k}_{y}^{(j)}
\tilde{\Lambda}^{(\text{B})}_{y}(i\omega_{n},i\Omega_{m}),~~~~~{\rm for}~j\in {\rm even},
\label{Vertex-Fun-B}
\end{eqnarray}
\end{subequations}
respectively, while the vertex kernels
$\tilde{\Lambda}^{(\text{A})}_{x}(i\omega_{n},i\Omega_{m})$,
$\tilde{\Lambda}^{(\text{A})}_{y}(i\omega_{n},i\Omega_{m})$,
$\tilde{\Lambda}^{(\text{B})}_{x}(i\omega_{n},i\Omega_{m})$,
and $\tilde{\Lambda}^{(\text{B})}_{y}(i\omega_{n},i\Omega_{m})$ satisfy the following
self-consistent equations,
\begin{subequations}\label{Lambda-Self-Consist}
\begin{eqnarray}
\hat{k}_{x}^{(j)}\tilde{\Lambda}_{x}^{(\text{A})}(i\omega_{n},i\Omega_{m})&+&\hat{k}_{y}^{(j)}
\tilde{\Lambda}_{y}^{(\text{A})}(i\omega_{n},i\Omega_{m})=n_{i}N\big\{
\sum_{\substack{\bm{k}\in\text{A}\\ j''\in {\rm odd}}}\tilde{T}_{jj''}(i\omega_{n}+i\Omega_{m})
\tilde{G}^{(\text{A})}_{\text{I}}(\bm{k},i\omega_{n}+i\Omega_{m})\nonumber\\
&\times&\big[\hat{k}^{(j'')}_{\rm F}\tau_{0}
+\hat{k}_{x}^{(j'')}\tilde{\Lambda}^{(\text{A})}_{x}(i\omega_{n},i\Omega_{m})
+\hat{k}_{y}^{(j'')}\tilde{\Lambda}^{(\text{A})}_{y}(i\omega_{n},i\Omega_{m})\big]
\tilde{G}^{(\text{A})}_{\text{I}}(\bm{k},i\omega_{n})\tilde{T}_{j''j}(i\omega_{n})
\nonumber\\
&+&\sum_{\substack{\bm{k}\in\text{B}\\ j''\in {\rm even}}}
\tilde{T}_{jj''}(i\omega_{n}+i\Omega_{m})
\tilde{G}^{(\text{B})}_{\text{I}}(\bm{k},i\omega_{n}+i\Omega_{m})\big[\hat{k}^{(j'')}_{\rm F}
\tau_{0}+\hat{k}_{x}^{(j'')}\tilde{\Lambda}^{(\text{B})}_{x}(i\omega_{n},i\Omega_{m})
\nonumber\\
&+& \hat{k}_{y}^{(j'')}\tilde{\Lambda}^{(\text{B})}_{y}(i\omega_{n},i\Omega_{m})\big]
\tilde{G}^{(\text{B})}_{\text{I}}(\bm{k},i\omega_{n})\tilde{T}_{j''j}(i\omega_{n})\big\},
~~~~~~~~~~{\rm for}~j\in {\rm odd}, \label{Lambda-Self-Consist-A}\\
\hat{k}_{x}^{(j)}\tilde{\Lambda}_{x}^{(\text{B})}(i\omega_{n},i\Omega_{m})&+&\hat{k}_{y}^{(j)}
\tilde{\Lambda}_{y}^{(\text{B})}(i\omega_{n},i\Omega_{m})=n_{i}N\big\{
\sum_{\substack{\bm{k}\in\text{A}\\ j''\in {\rm odd}}}
\tilde{T}_{j j''}(i\omega_{n}+i\Omega_{m})
\tilde{G}^{(\text{A})}_{\text{I}}(\bm{k},i\omega_{n}+i\Omega_{m})\nonumber\\
&\times& \big[\hat{k}^{(j'')}_{\rm F}\tau_{0}+\hat{k}_{x}^{(j'')}
\tilde{\Lambda}^{(\text{A})}_{x}(i\omega_{n},i\Omega_{m})
+\hat{k}_{y}^{(j'')}\tilde{\Lambda}^{(\text{A})}_{y}(i\omega_{n},i\Omega_{m})\big]
\tilde{G}^{(\text{A})}_{\text{I}}(\bm{k},i\omega_{n})\tilde{T}_{j''j}(i\omega_{n})
\nonumber\\
&+& \sum_{\substack{\bm{k}\in\text{B}\\ j''\in {\rm even}}}
\tilde{T}_{j j''}(i\omega_{n}+i\Omega_{m})
\tilde{G}^{(\text{B})}_{\text{I}}(\bm{k},i\omega_{n}+i\Omega_{m})
\big[\hat{k}^{(j'')}_{\rm F}\tau_{0}+\hat{k}_{x}^{(j'')}
\tilde{\Lambda}^{(\text{B})}_{x}(i\omega_{n},i\Omega_{m})\nonumber\\
&+&\hat{k}_{y}^{(j'')}\tilde{\Lambda}^{(\text{B})}_{y}(i\omega_{n},i\Omega_{m})\big]
\tilde{G}^{(\text{B})}_{\text{I}}(\bm{k},i\omega_{n})\tilde{T}_{j''j}(i\omega_{n})\big\},
~~~~~~~~~~{\rm for}~j\in {\rm even}. \label{Lambda-Self-Consist-B}
\end{eqnarray}
\end{subequations}
Substituting the above results in Eq. (\ref{Lambda-Self-Consist}) into
Eqs. (\ref{T-matrix-equation-1}) and  (\ref{corP-3}), the vertex-corrected electron
current-current correlation function (\ref{corP-3}) now can be expressed as,
\begin{eqnarray}\label{corP-4}
\mathop{\Pi}^{\leftrightarrow}(i\Omega_{m})&=&(eV^{\rm (TFA)}_{\rm F})^{2}{1\over N}
\sum\limits_{\bm{k}}{1\over\beta}\sum\limits_{i\omega_{n}}(\hat{k}_{x}+\hat{k}_{y})
Tr\big\{\tilde{G}_{\text{I}}(\bm{k},i\omega_{n})
\tilde{G}_{\text{I}}(\bm{k},i\omega_{n}+i\Omega_{m})[\hat{k}_{\rm F}\tau_{0}
+\hat{k}_{x}\tilde{\Lambda}_{x}(i\omega_{n},i\Omega_{m})
+\hat{k}_{y}\tilde{\Lambda}_{y}(i\omega_{n},i\Omega_{m})]\big\}\nonumber\\
&=& (eV^{\rm (TFA)}_{\rm F})^{2}\sum\limits_{j\in {\rm odd}}{1\over\beta}
\sum\limits_{i\omega_{n}}(\hat{k}_{x}^{(j)}+\hat{k}_{y}^{(j)})Tr\Big\{
{1\over N}\sum_{\bm{k}\in\text{A}}\tilde{G}^{(\text{A})}_{\text{I}}(\bm{k},i\omega_{n})
\tilde{G}^{(\text{A})}_{\text{I}}(\bm{k},i\omega_{n}+i\Omega_{m})[\hat{k}_{\rm F}^{(j)}\tau_{0}
+\hat{k}_{x}^{(j)}\tilde{\Lambda}_{x}^{(\text{A})}(i\omega_{n},i\Omega_{m})\nonumber\\
&+&\hat{k}_{y}^{(j)}\tilde{\Lambda}_{y}^{(\text{A})}(i\omega_{n},i\Omega)_{m}]\Big\}
+ (eV^{\rm (TFA)}_{\rm F})^{2}\sum\limits_{j\in {\rm even}}{1\over\beta}
\sum\limits_{i\omega_{n}}(\hat{k}_{x}^{(j)}+\hat{k}_{y}^{(j)})
Tr\Big\{{1\over N}\sum_{\bm{k} \in\text{B}}\tilde{G}^{(\text{B})}_{\text{I}}(\bm{k},i\omega_{n})
\tilde{G}^{(\text{B})}_{\text{I}}(\bm{k},i\omega_{n}+i\Omega_{m})\nonumber\\
&\times& [\hat{k}_{\rm F}^{(j)}\tau_{0}+\hat{k}_{x}^{(j)}
\tilde{\Lambda}_{x}^{(\text{B})}(i\omega_{n},i\Omega_{m})+\hat{k}_{y}^{(j)}
\tilde{\Lambda}_{y}^{(\text{B})}(i\omega_{n},i\Omega_{m})]\Big\}, ~~~~~~
\end{eqnarray}
with the electron Fermi velocity $V^{\rm (TFA)}_{\rm F}$ at around the tips of the Fermi arcs.
However, in the absence of an external magnetic field, the rotational symmetry in the system
is unbroken, indicating that $\Pi_{xy}(\Omega)=\Pi_{yx}(\Omega)=0$ and
$\Pi_{xx}(\Omega)=\Pi_{yy}(\Omega)$, and then the above vertex-corrected electron
current-current correlation function (\ref{corP-4}) is reduced as,
\begin{equation}\label{corP-5}
\mathop{\Pi}^{\leftrightarrow}(i\Omega_{m})=
\left(\begin{array}{cc}
\Pi_{xx}(i\Omega_{m}) & 0\\ 0 & \Pi_{yy}(i\Omega_{m})
\end{array}\right)
=\tau_{0}\Pi_{xx}(i\Omega_{m}),
\end{equation}
where $\Pi_{xx}(i\Omega_{m})$ is given by,
\begin{eqnarray}\label{corP-6}
\Pi_{xx}(i\Omega_{m})&=&(2eV^{\rm (TFA)}_{\rm F})^{2}{1\over\beta}\sum\limits_{i\omega_{n}}
J_{xx}(i\omega_{n},i\omega_{n}+i\Omega_{m}),
\end{eqnarray}
with the kernel function,
\begin{eqnarray}\label{current-density-2}
J_{xx}(i\omega_{n},i\omega_{n}+i\Omega_{m})&=&{1\over N}\sum_{\alpha=0}^{3}\Big\{\cos^{2}
\theta^{\rm (A)}_{\rm F}\tilde{I}^{(\text{A})}_{0}(\alpha,i\omega_{n},i\omega_{n}+i\Omega_{m})
Tr\big[\tau_{\alpha}[\tau_{0}+\tilde{\Lambda}_{x}^{(\text{A})}(i\omega_{n},i\Omega_{m})]\big]
\nonumber\\
&+&\cos^{2}\theta^{\rm (B)}_{\rm F}
\tilde{I}^{(\text{B})}_{0}(\alpha,i\omega_{n},i\omega_{n}+i\Omega_{m})
Tr\big[\tau_{\alpha}[\tau_{0}+\tilde{\Lambda}_{x}^{({B})}(i\omega_{n},i\Omega_{m})]\big]\Big\},
\end{eqnarray}
where the functions $\tilde{I}^{(\text{A})}_{0}(\alpha,i\omega_{n},i\omega_{n}+i\Omega_{m})$
and $\tilde{I}^{(\text{B})}_{0}(\alpha,i\omega_{n},i\omega_{n}+i\Omega_{m})$ are defined as,
\begin{subequations}\label{I-0-AB}
\begin{eqnarray}
\sum_{\bm{k}\in\text{A}}\tilde{G}^{\rm (A)}_{\rm I}(\bm{k},i\omega_{n})\tau_{\gamma}
\tilde{G}^{\rm (A)}_{\rm I}(\bm{k},i\omega_{n}+i\Omega_{m})
&=& \sum\limits_{\beta=0}^{3}
\tilde{I}_{\gamma}^{(\text{A})}(\beta,i\omega_{n},i\omega_{n}+i\Omega_{m})\tau_{\beta},~~\\
\sum_{\bm{k}\in\text{B}}\tilde{G}^{\rm (B)}_{\rm I}(\bm{k},i\omega_{n})\tau_{\gamma}
\tilde{G}^{\rm (B)}_{\rm I}(\bm{k},i\omega_{n}+i\Omega_{m})
&=& \sum\limits_{\beta=0}^{3}
\tilde{I}_{\gamma}^{(\text{B})}(\beta,i\omega_{n},i\omega_{n}+i\Omega_{m})\tau_{\beta},~~
\end{eqnarray}
\end{subequations}
\end{widetext}
respectively. After a quite complicated calculation, the function
$Tr[\tau_{\alpha}\tilde{\Lambda}^{(\text{A})}_{x}(\omega,\Omega)]$ in the above kernel function
(\ref{current-density-2}), which is a trace of the product of the vertex kernel
$\tilde{\Lambda}^{(\text{A})}_{x}(\omega,\Omega)$ and matrix $\tau_{\alpha}$ with
$\alpha=0,1,2,3$ in the region A of BZ, and the function
$Tr[\tau_{\alpha}\tilde{\Lambda}^{(\text{B})}_{x}(\omega,\Omega)]$ in the above kernel function
(\ref{current-density-2}), which is a trace of the product of the vertex kernel
$\tilde{\Lambda}^{(\text{B})}_{x}(\omega,\Omega)$ and matrix $\tau_{\alpha}$ in the region B of
BZ, can be derived straightforwardly [see Appendix \ref{vertex-correction}], and then the above
kernel function $J_{xx}(\omega,\omega+\Omega)$ can be obtained explicitly.

On the other hand, the dressed electron propagators $\tilde{G}_{\rm I}(\bm{k},i\omega_{n})$ and
$\tilde{G}_{\rm I}(\bm{k},i\omega_{n}+i\Omega_{m})$ are involved directly in the above kernel
function $J_{xx}(i\omega_{n},i\omega_{n}+i\Omega_{m})$ in Eq. (\ref{current-density-2}), then
the singularity of $J_{xx}(i\omega_{n},i\omega_{n}+i\Omega_{m})$ only lies at the real axes
[$\epsilon \in\mathbb{R}$] and these parallel to the real axes [$\epsilon-i\Omega_{m}$]. In
this case, the contribution for the summation of the kernel function
$J_{xx}(i\omega_{n},i\omega_{n}+i\Omega_{m})$ in Eq. (\ref{corP-6}) over the fermionic
Matsubara frequency $i\omega_{n}$ comes from the two branch cuts: $\epsilon \in\mathbb{R}$ and
$\epsilon-i\Omega_{m}$, and then the vertex-corrected electron current-current correlation
function (\ref{corP-6}) can be expressed as,
\begin{widetext}
\begin{eqnarray}\label{corP-7}
\Pi_{xx}(i\Omega_{m})&=& i(2eV^{\rm (TFA)}_{\rm F})^{2}\int_{-\infty}^{\infty}
{d\epsilon\over 2\pi}n_{\rm F}(\epsilon)\big[
J_{xx}(\epsilon+i\delta,\epsilon+i\Omega_{m})-J_{xx}(\epsilon-i\delta,\epsilon+i\Omega_{m})
\nonumber\\
&+& J_{xx}(\epsilon-i\Omega_{m},\epsilon+i\delta)
-J_{xx}(\epsilon-i\Omega_{m},\epsilon-i\delta)\big],
\end{eqnarray}
By virtue of the analytical continuation $i\Omega_{m}\to\Omega+i\delta$, the above
vertex-corrected electron current-current correlation function (\ref{corP-7}) can be
obtained explicitly as,
\begin{eqnarray}\label{corP-8}
\Pi_{xx}(\Omega) &=& i(2eV^{\rm (TFA)}_{\rm F})^{2}\int_{-\infty}^{\infty}{d\epsilon\over 2\pi}
\Big\{ n_{\rm F}(\epsilon)\big[ J_{xx}(\epsilon+i\delta,\epsilon+\Omega+i\delta)
-J_{xx}(\epsilon-i\delta,\epsilon+\Omega+i\delta)\big] \nonumber\\
&+& n_{\rm F}(\epsilon+\Omega)\big[ J_{xx}(\epsilon-i\delta,\epsilon+\Omega+i\delta)
- J_{xx}(\epsilon-i\delta,\epsilon+\Omega-i\delta)\big]\Big\},
\end{eqnarray}
and then the microwave conductivity
$\mathop{\sigma}\limits^{\leftrightarrow}(\Omega,T)=\tau_{0}\sigma(\Omega,T)$
in Eq. (\ref{conductivity-1}) in the presence of impurities is obtained as,
\begin{eqnarray}\label{Conductivity-2}
\sigma(\Omega)=-{{\rm Im}\Pi_{xx}(\Omega)\over\Omega}
=(2eV^{\rm (TFA)}_{\rm F})^{2}\int_{-\infty}^{\infty}{d\epsilon\over 2\pi} {n_{\rm F}(\epsilon)
-n_{\rm F}(\epsilon+\Omega)\over\Omega}[{\rm Re}J_{xx}(\epsilon-i\delta,\epsilon+\Omega+i\delta)
- Re J_{xx}(\epsilon+i\delta,\epsilon+\Omega+i\delta)].\nonumber\\
\end{eqnarray}
\end{widetext}

\section{Quantitative characteristics}\label{Quantitative-characteristics}

\begin{figure}
\includegraphics[scale=0.35]{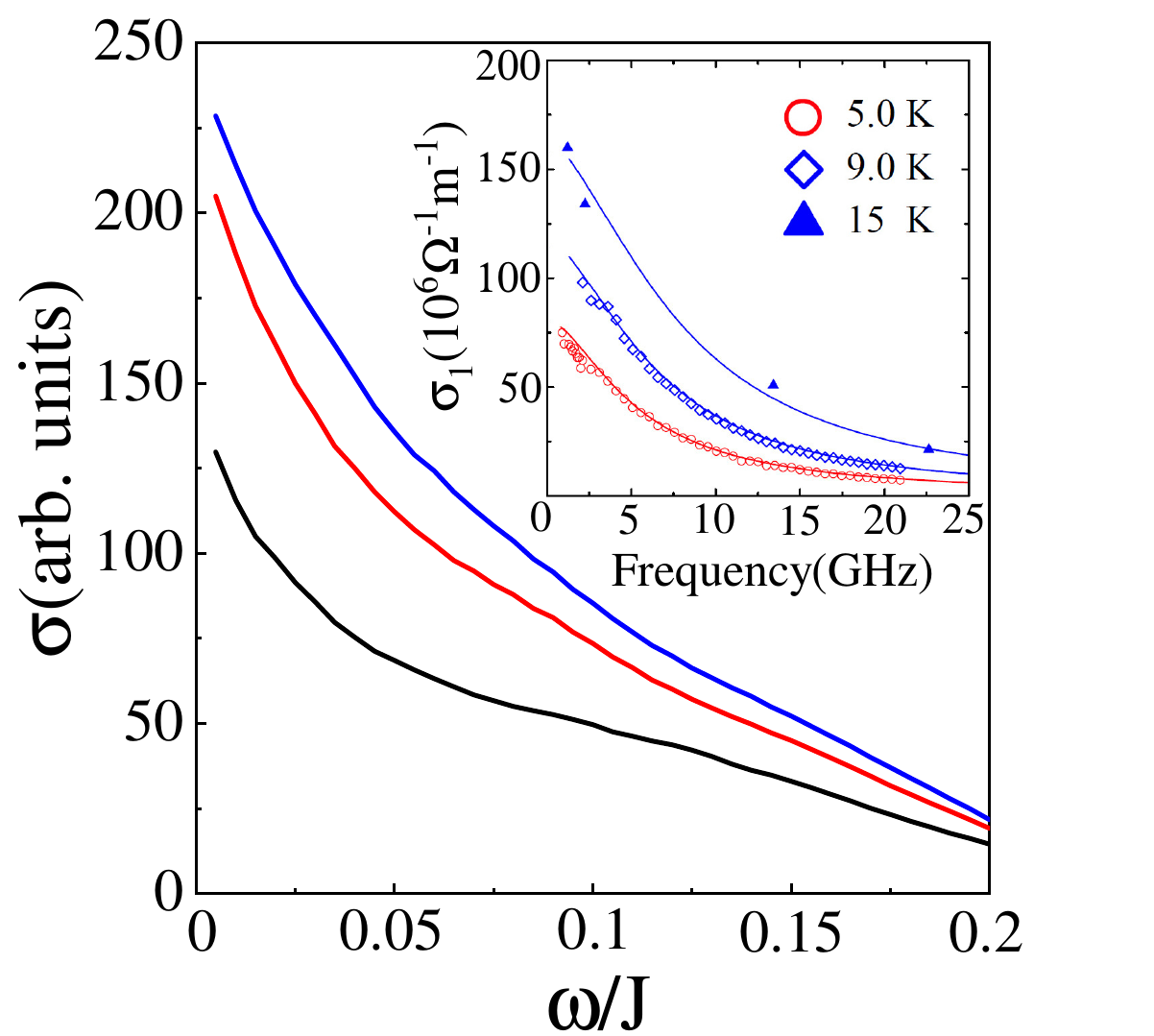}
\caption{(Color online) The microwave conductivity as a function of energy at the doping
concentration $\delta=0.15$ for temperatures $T=0.005J\sim 5$K (black-line), $T=0.009J\sim 9$K
(red-line), and $T=0.015J\sim 15$K (blue-line) together with the impurity concentration
$n_{i}=0.0025$ and parameter of the impurity scattering potential strength $d=0.05$. Inset:
the corresponding experimental result of the microwave conductivity observed on
YBa$_{2}$Cu$_{3}$O$_{6.993}$ taken from Ref. \onlinecite{Harris06}.
\label{conductivity-spectrum-energy}}
\end{figure}

In the self-consistent $T$-matrix approach, the strength of the impurity scattering potential
is an important parameter. Unless otherwise indicated, the adjacent-tip impurity scattering
$V_{2}$, $V_{3}$, $V_{7}$, and $V_{8}$, and the opposite-tip impurity scattering $V_{4}$,
$V_{5}$, and $V_{6}$ in the following discussions are chosen as $V_{2}=0.85V_{1}$,
$V_{3}=0.8V_{1}$, $V_{7}=0.8V_{1}$, $V_{8}=0.9V_{1}$, $V_{4}=0.7V_{1}$, $V_{5}=0.65V_{1}$,
and $V_{6}=0.75V_{1}$, respectively, as in the previous discussions of the influence of the
impurity scattering on the electronic structure\cite{Zeng22}, while the strength of the
intra-tip impurity scattering $V_{1}$ is chosen as
$V_{1}=V_{\rm scale}{\rm tan}({\pi\over 2}d)$ with $V_{\rm scale}=58J$ and the adjustable
parameter $d$ of the impurity scattering potential strength, where the case of $d\sim 0$
[then ${\rm tan}({\pi\over 2}d)\sim 0$] is corresponding to the case $V_{j}\sim 0$ with
$j=1,2,3,...8$ in
the Born-limit, while the case of $d\sim 1$ [then ${\rm tan}({\pi\over 2}d)\sim \infty$] is
corresponding to the case $V_{j}\sim\infty$ in the unitary-limit.

\begin{figure}
\includegraphics[scale=0.35]{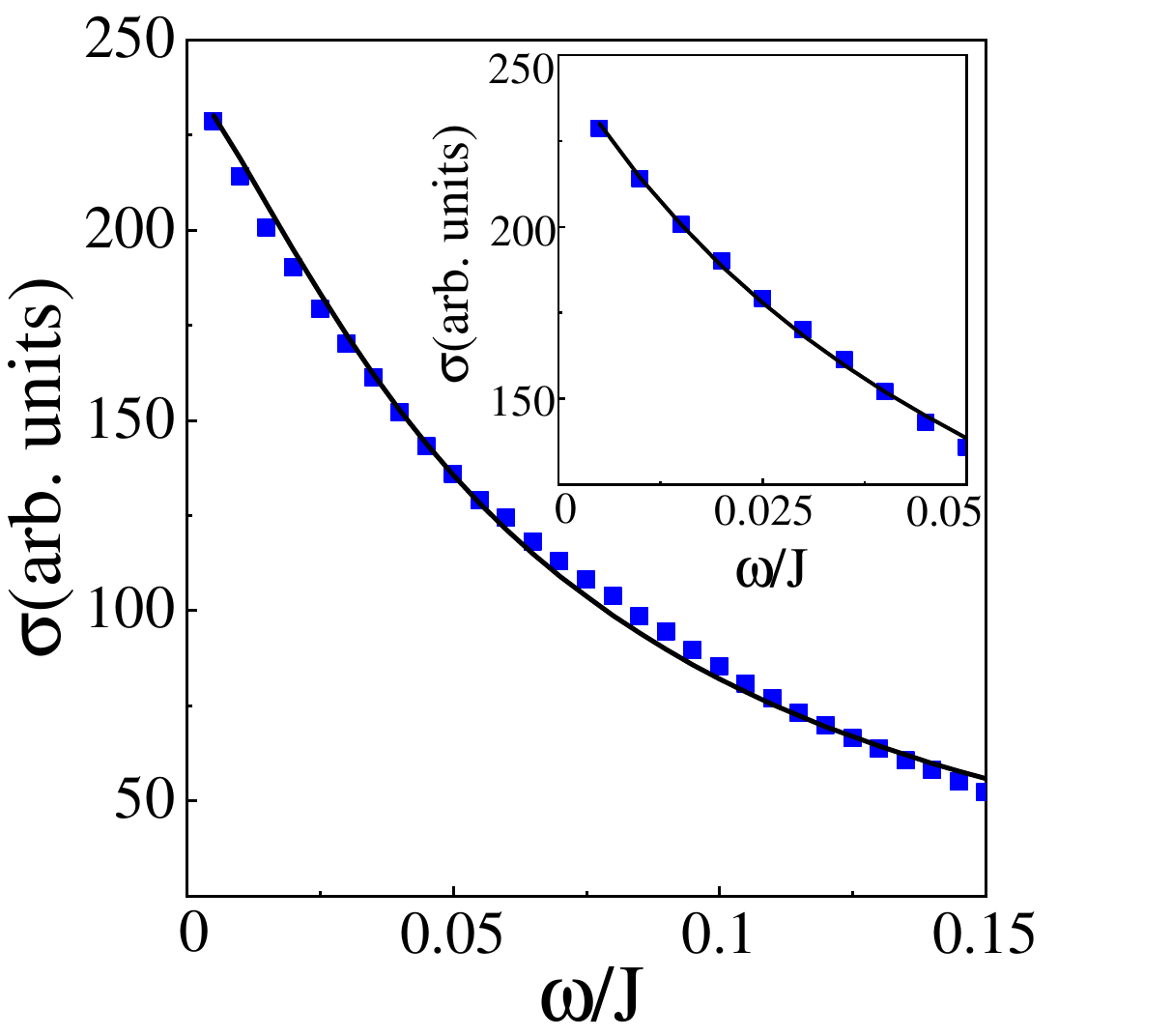}
\caption{(Color online) The numerical fit (black-line) with Eq. (\ref{non-Drude-form}).
The blue-squares are the result of the microwave conductivity with $T=0.015J\sim 15$K taken
from Fig. \ref{conductivity-spectrum-energy}. Inset: The numerical fit (black-line) with
the fit form $\sigma(\omega,T)=A_{0}/[\omega+B_{0}]$, where $A_{0}=15.676$ and
$B_{0}=0.063$. The blue-squares are the result of the low-energy microwave conductivity
with $T=0.015J\sim 15$K taken from Fig. \ref{conductivity-spectrum-energy}.
\label{conductivity-spectrum-fit}}
\end{figure}

We are now ready to discuss the effect of the impurity scattering on the microwave conductivity
in cuprate superconductors. We have performed a calculation for the microwave conductivity
$\sigma(\omega,T)$ in Eq. (\ref{Conductivity-2}), and the results of the microwave conductivity
$\sigma(\omega,T)$ as a function of energy at the doping concentration $\delta=0.15$ for
temperatures $T=0.005J\sim 5$K
(black-line), $T=0.009J\sim 9$K (red-line), and $T=0.015J\sim 15$K (blue-line) together with
the impurity concentration $n_{i}=0.0025$ and parameter of the impurity scattering potential
strength $d=0.05$ are plotted in Fig. \ref{conductivity-spectrum-energy} in comparison with the
corresponding experimental results of the microwave conductivity observed on the cuprate
superconductor\cite{Harris06} YBa$_{2}$Cu$_{3}$O$_{6.993}$ (inset). The results in
Fig. \ref{conductivity-spectrum-energy} therefore show clearly that the energy dependence of the
low-temperature microwave conductivity in cuprate superconductor
\cite{Bonn93,Lee96,Hosseini99,Turner03,Harris06} is qualitatively reproduced, where the highly
unconventional features of the low-temperature microwave conductivity spectrum can be summarized
as: (i) a sharp cusp-like peak develops at the low-energy limit; (ii) the low-temperature
microwave conductivity spectrum is non-Drude-like; (iii) a high-energy tail falls slowly with
the increase of energy. To see this non-Drude behavior in the low-temperature microwave
conductivity spectrum more clearly, the results of the low-temperature microwave conductivity
spectra shown in Fig. \ref{conductivity-spectrum-energy} have been numerically fitted in terms
of the following fit form,
\begin{eqnarray}\label{non-Drude-form}
\sigma(\omega,T)={\sigma_{0}\over 1+(\omega/C_{0}T)^{\rm y}},
\end{eqnarray}
as they have been done in the experiments\cite{Turner03}, and the fit result at the temperature
$T=0.015J\sim 15$K is plotted in Fig. \ref{conductivity-spectrum-fit} (black-line), where the
fit parameters $\sigma_{0}=238.073$, $C_{0}=4.145$, and ${\rm y}=1.333$. For
a more better understanding, we have also fitted the low-energy part of the microwave
conductivity spectrum alone with the fit form $\sigma(\omega,T)=A_{0}/[\omega+B_{0}]$, and
the numerically fit result at the same temperature $T=0.015J\sim 15$K is also plotted in
Fig. \ref{conductivity-spectrum-fit} (inset), where the fit parameters $A_{0}=15.676$ and
$B_{0}=0.063$. These fit results in Fig. \ref{conductivity-spectrum-fit} thus indicate
clearly that although the lower-energy cusp-like peak in Fig. \ref{conductivity-spectrum-energy}
decay as $\rightarrow 1/[\omega+B_{0}]$, the overall shape of the low-temperature microwave
conductivity spectrum in Fig. \ref{conductivity-spectrum-energy} exhibits a special
non-Drude-like behavior, which can be well fitted by the formula in Eq. (\ref{non-Drude-form}),
in agreement with the corresponding experimental observations\cite{Turner03,Harris06}. More
specifically, in comparison with other fit results at the temperatures $T=0.005J\sim 5$K and
$T=0.009J\sim 9$K, we also find that the fit parameter ${\rm y}$ in the fit form
(\ref{non-Drude-form}) is almost independence of temperature, and remains relatively constant,
taking the average value of ${\rm y}=1.333$. This anticipated value of the fit parameter
${\rm y}=1.333$ is not too far from the corresponding value of ${\rm y}=1.45(\pm 0.06)$, which
has been employed in Ref. \onlinecite{Turner03} to fit the corresponding experimental data with
the same fit formula (\ref{non-Drude-form}). The qualitative agreement between the present
theoretical results and experimental data therefore also show that the kinetic-energy-driven
superconductivity, incorporating the effect of the impurity scattering within the framework of
the self-consistent $T$-matrix theory, can give a consistent description of the low-temperature
microwave conductivity spectrum found in the microwave surface impedance measurements on cuprate
superconductors\cite{Bonn93,Lee96,Hosseini99,Turner03,Harris06}.

\begin{figure}
\includegraphics[scale=0.3]{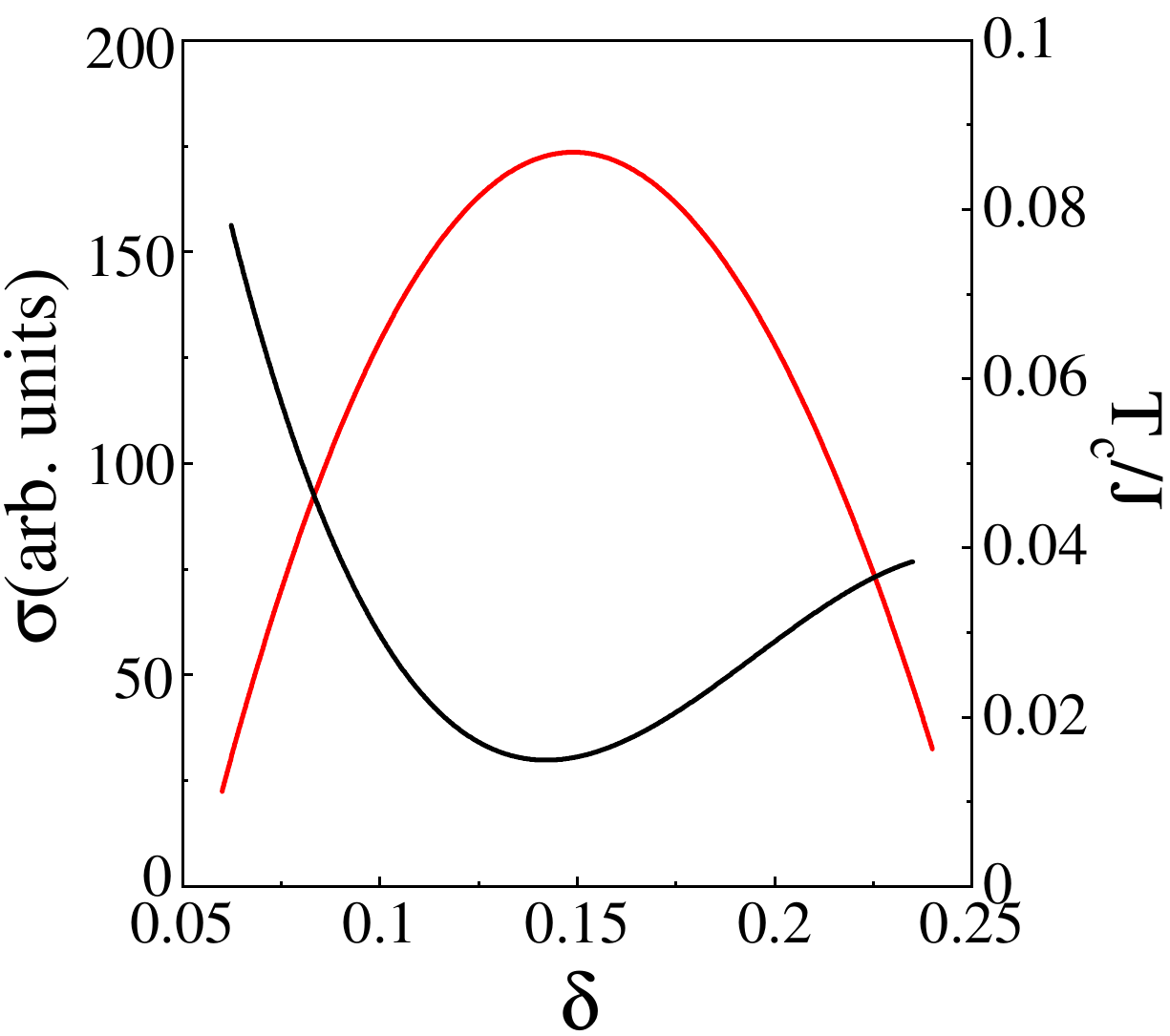}
\caption{(Color online) The microwave conductivity (black-line) as a function of doping with
$T=0.002J$ for $\omega=0.0025J$ together with $n_{i}=0.0025$ and $d=0.05$. The red-line is the
corresponding result of $T_{\rm c}$. \label{conductivity-spectrum-doping}}
\end{figure}

As a natural consequence of the doped Mott insulator, the microwave conductivity in cuprate
superconductors evolve with doping. In Fig. \ref{conductivity-spectrum-doping}, we plot the
result of $\sigma(\omega,T)$ [black-line] as a function of doping with $T=0.002J$
for energy $\omega=0.0025J$ together with $n_{i}=0.0025$ and $d=0.05$. For a comparison,
the corresponding result\cite{Feng12,Feng15,Feng15a} of $T_{\rm c}$ obtained within the
framework of the kinetic-energy-driven superconductivity is also shown in
Fig. \ref{conductivity-spectrum-doping} (red-line). Apparently, in a striking contrast to the
dome-like shape of the doping dependence of $T_{\rm c}$, the microwave conductivity exhibits
a reverse dome-like shape of the doping dependence, where $\sigma(\omega,T)$ is a decreasing
function of the doping concentration, the system is thought to be at the underdoped regime.
The system is at around the {\it optimal doping}, where $\sigma(\omega,T)$ reaches its minimum.
\begin{figure}[h!]
\centering
\includegraphics[scale=0.3]{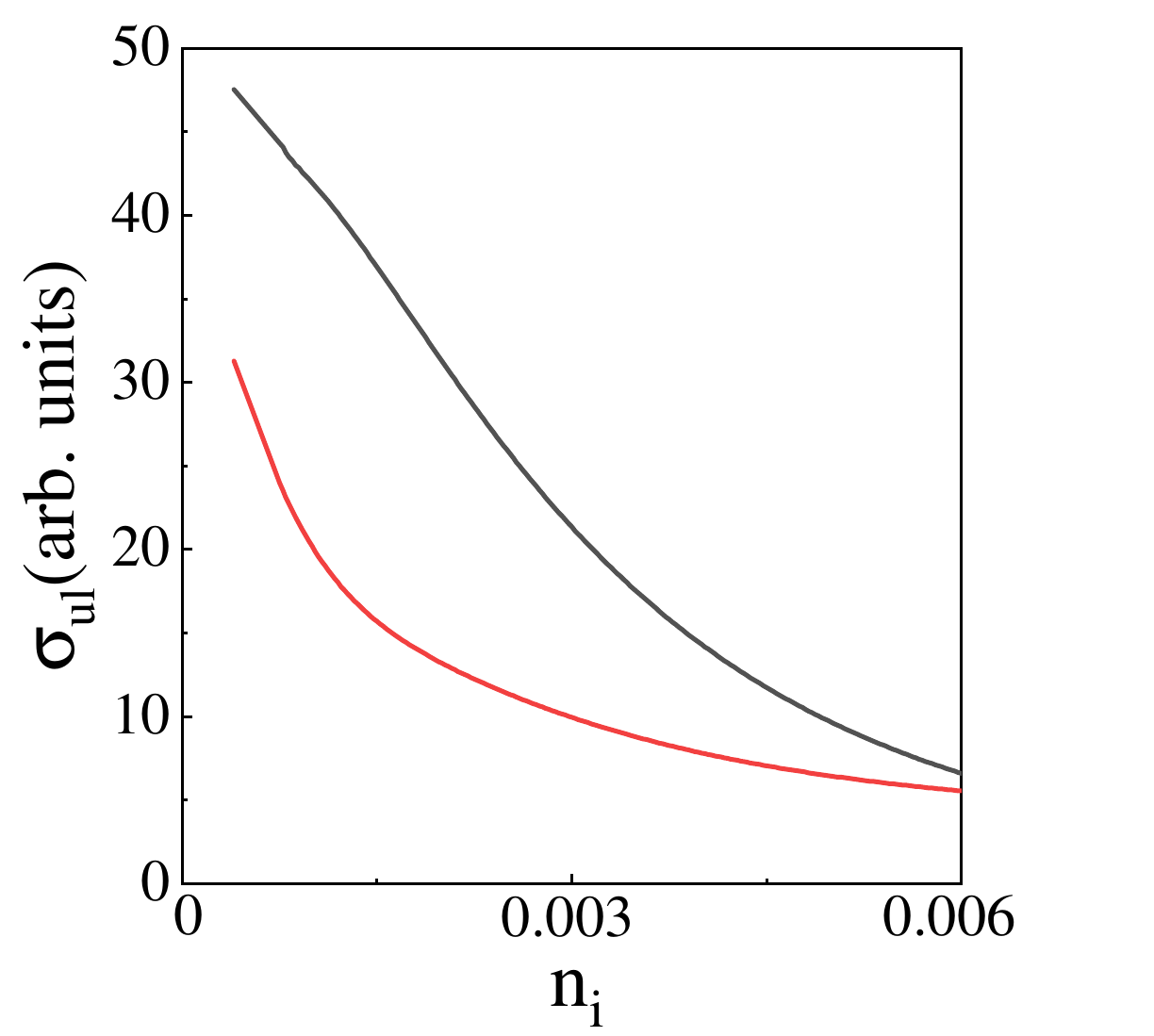}
\caption{(Color online) The microscopic conductivity in the universal-limit as a function of
the impurity concentration at $\delta=0.15$ with $T=0.002J$ for $d=0.05$ (black-line) and
$d=0.5$ (red-line).\label{conductivity-spectrum-impurity}}
\end{figure}
However, with the further increase in the doping concentration, $\sigma(\omega,T)$ increases at
the overdoped regime. This reverse dome-like shape of the doping dependence of the microwave
conductivity in low energies and low temperatures is also qualitatively consistent with the
microwave conductivity $\sigma_{\rm ul}\propto 1/\bar{\Delta}$ in the universal limit of
$\omega\rightarrow 0$ and $T\rightarrow 0$, since the SC gap parameter $\bar{\Delta}$ obtained
within the framework of the kinetic-energy-driven superconductivity\cite{Feng12,Feng15,Feng15a}
has the similar dome-like shape of the doping dependence.

For a further understanding of the intrinsic effect of the impurity scattering on the SC-state
quasiparticle transport in cuprate superconductors, we now turn to discuss the evolution of the
microwave conductivity with the impurity concentration in the case of the universal-limit. The
microwave conductivity $\sigma_{\rm ul}$ in the universal-limit can be obtained directly from
the energy and temperature dependence of the microwave conductivity (\ref{Conductivity-2}) in
the zero-temperature ($T\to 0$) and zero-energy ($\Omega\to 0$) limits as,
\begin{eqnarray}\label{Conductivity-3}
\sigma_{\rm ul}&=&\lim_{\substack{\Omega\to 0\\ T\to 0}}\sigma_{xz}(\Omega)\nonumber\\
&=&{(2eV^{\rm (TFA)}_{\rm F})^{2}\over 2\pi}
\lim_{\epsilon\to 0}[{\rm Re}J_{xx}(\epsilon-i\delta,\epsilon+i\delta)\nonumber\\
&-& {\rm Re}J_{xx}(\epsilon+i\delta,\epsilon+i\delta)].~~~~~~
\end{eqnarray}
In this case, we have made a series of calculations for $\sigma_{\rm ul}$ at different impurity
concentrations and different strengths of the impurity scattering potential, and the results of
$\sigma_{\rm ul}$ as a function of the impurity concentration $n_{i}$ at $\delta=0.15$ for
$d=0.05$ (black-line) and $d=0.5$ (red-line) are plotted in
Fig. \ref{conductivity-spectrum-impurity}, where the main features can be summarized as: (i) for
a given set of the impurity scattering potential strength, the microwave conductivity gradually
decreases with the increase of the impurity concentration; (ii) for a given impurity
concentration, the microwave conductivity decreases when the strength of the impurity scattering
potential is increased. In other words, the crucial role played by the impurity scattering is
the further reduction of the microwave conductivity.

\begin{figure}[h!]
\centering
\includegraphics[scale=0.3]{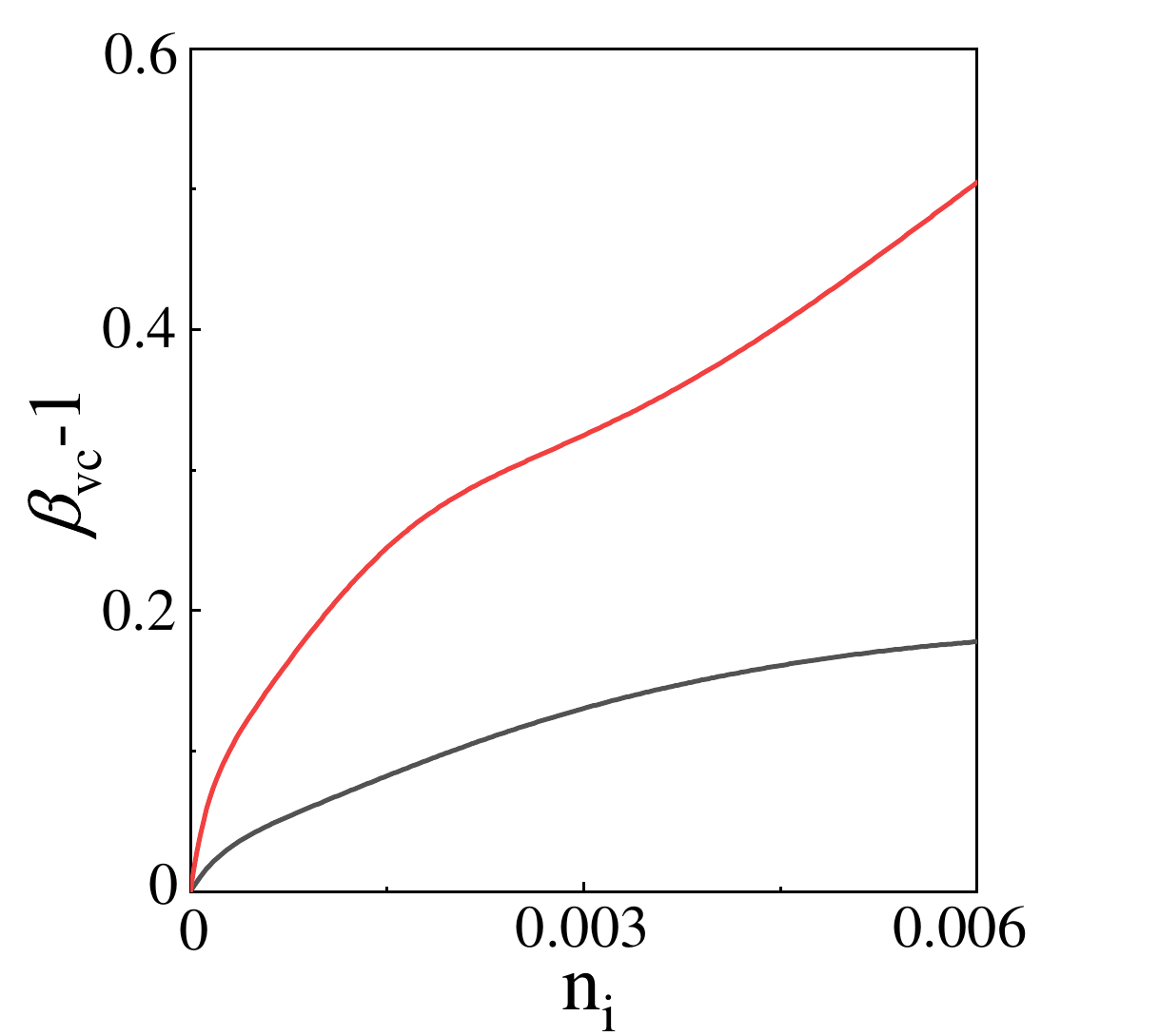}
\caption{(Color online) The characteristic factor of the impurity-induced vertex correction as
a function of the impurity concentration at $\delta=0.15$ for $d=0.05$ (black-line) and
$d=0.5$ (red-line). \label{factor-impurity}}
\end{figure}

In the present theoretical framework, the effect of the strong electron correlation on the
microwave conductivity is reflected in the homogenous part of the electron propagator (then
the homogenous self-energy), while the effect of the impurity scattering on the microwave
conductivity is reflected both in the impurity-dressed electron propagator (then the
impurity-scattering self-energy) and the impurity-induced vertex correction to the electron
current-current correlation function. In other words, the microwave conductivity is further
renormalized by the impurity-induced vertex correction. For the understanding of this
renormalization of the microwave conductivity from the impurity-induced vertex correction,
the microwave conductivity in the case of the universal-limit in Eq. (\ref{Conductivity-3})
can be rewritten as,
\begin{equation}\label{correction-factor}
\sigma_{\rm ul}=\beta_{\text{vc}}\sigma_{\rm ul}^{(0)},
\end{equation}
where the characteristic factor $\beta_{\text{vc}}$ is the impurity-induced vertex correction
to the universal bare result of the microscopic conductivity $\sigma_{\rm ul}^{(0)}$, while
this $\sigma_{\rm ul}^{(0)}$ can be reduced directly from $\sigma_{\rm ul}$ in
Eq. (\ref{Conductivity-3}) by ignoring the impurity-induced vertex correction as,
\begin{widetext}
\begin{eqnarray}\label{Bare-Conduct}
\sigma_{\rm ul}^{(0)} &=& {(2eV^{\rm (TFA)}_{\rm F})^{2}\over \pi}
\lim_{\epsilon \to 0}\sum_{\mu=\text{A},\text{B}}\Theta^{(\mu)}(\theta_{\rm F}){\rm Re}
\big [ \tilde{I}^{(\mu)}_{0}(0,\epsilon-i\delta,\epsilon+i\delta)
-\tilde{I}^{(\mu)}_{0}(0,\epsilon+i\delta,\epsilon+i\delta)\big ],~~~~
\end{eqnarray}
\end{widetext}
with the function,
\begin{eqnarray}
\Theta^{(\mu)}(\theta_{\rm F})= \left \{
\begin{array}{lr}
\cos\theta^{\rm (A)}_{\rm F},~~~{\rm for}\ \mu=A\\
\cos\theta^{\rm (B)}_{\rm F},~~~{\rm for}\ \mu=B
\end{array}\right.
\end{eqnarray}
In Fig. \ref{factor-impurity}, we plot characteristic factor $\beta_{\text{vc}}-1$ as a
function of the impurity concentration $n_{i}$ at $\delta=0.15$ for $d=0.05$ (black-line)
and $d=0.5$ (red-line), where for a given set of the impurity scattering potential strength,
the characteristic factor monotonically {\it increases} as the impurity concentration is
increased. On the other hand, for a given impurity concentration, $\beta_{\text{vc}}-1$
{\it increases} with the increase of the strength of the impurity scattering potential. It
thus shows clearly that the impurity-induced vertex correction is quite significant in the
renormalization of the microwave conductivity
\cite{Durst00,Berlinsky00,Hettler00,Durst02,Kim04,Nunner05,Wang08},
and then all the effects of the strong electron correlation, the impurity-scattering
self-energy, and the impurity-induced vertex correction lead to the highly unconventional
behaviors in the microwave conductivity of cuprate
superconductors\cite{Bonn93,Lee96,Hosseini99,Turner03,Harris06}.

\section{Summary}\label{conclude}

Starting from the homogenous electron propagator and the related microscopic octet scattering
model, which are obtained within the framework of the kinetic-energy-driven superconductivity,
we have rederived the impurity-dressed electron propagator in the self-consistent $T$-matrix
approach, where the impurity scattering self-energy is evaluated in the Fermi-arc-tip
approximation of the quasiparticle excitations and scattering processes, and then the
impurity-dressed electron propagator incorporates both the strong electron correlation and
impurity-scattering effects. By virtue of this impurity-dressed electron propagator, we then
have investigated the effect of the impurity scattering on the low-temperature microwave
conductivity of cuprate superconductors, where the electron current-current correlation
function is derived by taking into account the impurity-induced vertex correction. The
obtained results show clearly that the low-temperature microwave conductivity spectrum is a
non-Drude-like, with a sharp cusp-like peak extending to zero-energy and a high-energy tail
falling slowly with energy, in agreement with the corresponding experimental observations
\cite{Bonn93,Lee96,Hosseini99,Turner03,Harris06}. In particular, although the low-energy
cusp-like peak decay as $\rightarrow A_{0}/[\omega+B_{0}]$, the overall shape of the
low-temperature microwave conductivity spectrum exhibits a special non-Drude-like behavior,
and can be well fitted by the formula $\sigma(\omega,T)=\sigma_{0}/[1+(\omega/C_{0}T)^{\rm y}]$
with the relatively temperature-independent constant ${\rm y}$. Moreover, the low-temperature
microwave conductivity decreases with the increase of the impurity concentration or with the
increase of the strength of the impurity scattering potential. Our results therefore indicate
that the highly unconventional features of the microwave conductivity in cuprate
superconductors are arisen from both the strong electron correlation and impurity-scattering
effects. The theory also predicts a reverse dome-like shape of the doping dependence of the
microwave conductivity, which is in a striking contrast to the dome-like shape of the doping
dependence of $T_{\rm c}$, and therefore should be verified by further experiments.

\section*{Acknowledgements}

This work is supported by the National Key Research and Development Program of China under
Grant No. 2021YFA1401803, and the National Natural Science Foundation of China under Grant
Nos. 12247116, 11974051, and 12274036.


\begin{widetext}

\begin{appendix}

\section{Derivation of vertex kernels of electron current-current correlation function}
\label{vertex-correction}

Starting from the homogenous part of the d-wave BCS type formalism, the electron
current-current correlation function has been discussed by taking into account the
impurity-induced vertex correction
\cite{Durst00,Berlinsky00,Hettler00,Durst02,Kim04,Nunner05,Wang08}, where the
$T$-matrix approach has been employed to derive the vertex kernels of the electron
current-current correlation function in the nodal approximation. In this
Appendix \ref{vertex-correction}, we generalize these previous calculations
\cite{Durst00,Berlinsky00,Hettler00,Durst02,Kim04,Nunner05,Wang08} for the vertex kernels of
the electron current-current correlation function in the nodal approximation to the present
case in the Fermi-arc-tip approximation. In the microscopic octet scattering model shown in
Fig. \ref{Fermi-arc-tip-picture}, the tips of the Fermi arcs labelled by the odd numbers are
located in the region A of BZ, where $|k_{y}|>|k_{x}|$, while the tips of the Fermi arcs
labelled by the even numbers are located in the region B of BZ, where $|k_{x}|>|k_{y}|$. For a
convenience in the following discussions, $j=1$ in Eq. (\ref{Lambda-Self-Consist-A}) is chosen
in the region A of BZ, and $j=2$ in Eq. (\ref{Lambda-Self-Consist-B}) is chosen in the region
B of BZ, then the trace of the product between the self-consistent equation
(\ref{Lambda-Self-Consist-A}) and the unit vector $\hat{k}_{x}^{(1)}$ in the region A and the
trace of the product between the self-consistent equation (\ref{Lambda-Self-Consist-B}) and
the unit vector $\hat{k}_{x}^{(2)}$ in the region B can be obtained as,
\begin{subequations}\label{Itr-Lambda}
\begin{eqnarray}
&&Tr[\tau_{0}\tilde{\Lambda}^{(\text{A})}_{x}(\omega,\Omega)]={n_{i}N\over\cos^{2}
\theta^{\rm (A)}_{\rm F}}
\sum_{\bm{k}\in\text{A}}Tr\big[\tilde{G}^{(\text{A})}_{\text{I}}(\bm{k},\omega)
\sum_{j''\in {\rm odd}}\hat{k}_{x}^{(1)}\cdot\hat{k}^{(j'')}_{\rm F}\tilde{T}_{j''1}(\omega)
\tilde{T}_{1j''}(\omega+\Omega)\tilde{G}^{(\text{A})}_{\text{I}}(\bm{k},\omega+\Omega)
[\tau_{0}+\tilde{\Lambda}^{(\text{A})}_{x}(\omega,\Omega)]\big] \nonumber\\
&&+ {n_{i}N\over\cos^{2}\theta^{\rm (A)}_{\rm F}}\sum_{\bm{k}\in\text{B}}
Tr\big[\tilde{G}^{(\text{B})}_{\text{I}}(\bm{k},\omega)\sum_{j''\in {\rm even}}
\hat{k}_{x}^{(1)}\cdot\hat{k}^{(j'')}_{\rm F}\tilde{T}_{j''1}(\omega)
\tilde{T}_{1j''}(\omega+\Omega)\tilde{G}^{(\text{B})}_{\text{I}}(\bm{k},\omega+\Omega)[\tau_{0}
+\tilde{\Lambda}^{(\text{B})}_{x}(\omega,\Omega)]\big],~~~~\label{Itr-Lambda-o}\\
&&Tr[\tau_{0}\tilde{\Lambda}^{(\text{B})}_{x}(\omega,\Omega)]={n_{i}N\over\cos^{2}
\theta^{\rm (B)}_{\rm F}}
\sum_{\bm{k}\in\text{A}}Tr\big[\tilde{G}^{(\text{A})}_{\text{I}}(\bm{k},\omega)
\sum_{j''\in {\rm odd}}\hat{k}_{x}^{(2)}\cdot\hat{k}^{(j'')}_{\rm F}\tilde{T}_{j''2}(\omega)
\tilde{T}_{2 j''}(\omega+\Omega)\tilde{G}^{(\text{A})}_{\text{I}}(\bm{k},\omega+\Omega)
[\tau_{0}+\tilde{\Lambda}^{(\text{A})}_{x}(\omega,\Omega)]\big] \nonumber\\
&&+{n_{i}N\over\cos^{2}\theta^{\rm (B)}_{\rm F}}\sum_{\bm{k}\in\text{B}}
Tr\big[\tilde{G}^{(\text{B})}_{\text{I}}(\bm{k},\omega)\sum_{j''\in {\rm even}}\hat{k}_{x}^{(2)}
\cdot\hat{k}^{(j'')}_{\rm F}\tilde{T}_{j''2}(\omega)\tilde{T}_{2j''}(\omega+\Omega)
\tilde{G}^{(\text{B})}_{\text{I}}(\bm{k},\omega+\Omega)[\tau_{0}
+\tilde{\Lambda}^{(\text{B})}_{x}(\omega,\Omega)]\big],~~~~~~~~\label{Itr-Lambda-e}
\end{eqnarray}
\end{subequations}
respectively, where the Fermi velocity unit vectors $\hat{k}_{\rm F}^{(j)}$ with $j=1,2,3,...,8$
at the tips of the Fermi-arc are defined as follows:
$\hat{k}_{\rm F}^{(1)}=\hat{k}_{x}\cos\theta_{\rm F}+\hat{k}_{y}\sin\theta_{\rm F}$,
$\hat{k}_{\rm F}^{(2)}=\hat{k}_{x}\sin\theta_{\rm F}+\hat{k}_{y}\cos\theta_{\rm F}$,
$\hat{k}_{\rm F}^{(3)}=\hat{k}_{x}\cos\theta_{\rm F}-\hat{k}_{y}\sin\theta_{\rm F}$,
$\hat{k}_{\rm F}^{(4)}=\hat{k}_{x}\sin\theta_{\rm F}-\hat{k}_{y}\cos\theta_{\rm F}$,
$\hat{k}_{\rm F}^{(5)}=-\hat{k}_{x}\cos\theta_{\rm F}-\hat{k}_{y}\sin\theta_{\rm F}$,
$\hat{k}_{\rm F}^{(6)}=-\hat{k}_{x}\sin\theta_{\rm F}-\hat{k}_{y}\cos\theta_{\rm F}$,
$\hat{k}_{\rm F}^{(7)}=-\hat{k}_{x}\cos\theta_{\rm F}+\hat{k}_{y}\sin\theta_{\rm F}$,
$\hat{k}_{\rm F}^{(8)}=-\hat{k}_{x}\sin\theta_{\rm F}+\hat{k}_{y}\cos\theta_{\rm F}$.
In particular, it is easy to verify the following relations,
\begin{subequations}\label{To-sum-o}
\begin{eqnarray}
{n_{i}N\over\cos^{2}\theta_{\rm F}}\sum_{j''\in {\rm odd}}\hat{k}_{x}^{(1)}\cdot
\hat{k}_{\rm F}^{(j'')}\tilde{T}_{j''1}(\omega)\tilde{T}_{1j''}(\omega+\Omega)
&=& n_{i}N\Big[\tilde{T}_{11}(\omega)\tilde{T}_{11}(\omega+\Omega)+\tilde{T}_{31}(\omega)
\tilde{T}_{13}(\omega+\Omega) \nonumber\\
&-& \tilde{T}_{51}(\omega)\tilde{T}_{15}(\omega+\Omega)-\tilde{T}_{71}(\omega)
\tilde{T}_{17}(\omega+\Omega)\Big],~~~~\\
{n_{i}N\over\cos^{2}\theta_{\rm F}}\sum_{j''\in {\rm even}}\hat{k}_{x}^{(1)}\cdot
\hat{k}_{\rm F}^{(j'')}\tilde{T}_{j''1}(\omega)\tilde{T}_{1j''}(\omega+\Omega)
&=& \tan \theta_{\rm{F}} n_{i}N\Big[\tilde{T}_{21}(\omega)\tilde{T}_{12}(\omega+\Omega)
+\tilde{T}_{41}(\omega)\tilde{T}_{14}(\omega+\Omega) \nonumber\\
&-& \tilde{T}_{61}(\omega)\tilde{T}_{16}(\omega+\Omega)-\tilde{T}_{81}(\omega)
\tilde{T}_{18}(\omega+\Omega)\Big],~~~~\\
{n_{i}N\over\sin^{2}\theta_{\rm F}}\sum_{j''\in {\rm odd}}\hat{k}_{x}^{(2)}\cdot
\hat{k}_{\rm F}^{(j'')}\tilde{T}_{j''2}(\omega)\tilde{T}_{2j''}(\omega+\Omega)
&=& \cot \theta_{\rm{F}} n_{i}N\Big[\tilde{T}_{12}(\omega)\tilde{T}_{21}(\omega+\Omega)
+\tilde{T}_{32}(\omega)\tilde{T}_{23}(\omega+\Omega) \nonumber\\
&-& \tilde{T}_{52}(\omega)\tilde{T}_{25}(\omega+\Omega)-\tilde{T}_{72}(\omega)
\tilde{T}_{27}(\omega+\Omega)\Big],~~~~
\end{eqnarray}
\begin{eqnarray}
{n_{i}N\over\sin^{2}\theta_{\rm F}}\sum_{j''\in {\rm even}}\hat{k}_{x}^{(2)}\cdot
\hat{k}_{\rm F}^{(j'')}\tilde{T}_{j''2}(\omega)\tilde{T}_{2j''}(\omega+\Omega)
&=& n_{i}N\Big[\tilde{T}_{22}(\omega)\tilde{T}_{22}(\omega+\Omega)+\tilde{T}_{42}(\omega)
\tilde{T}_{24}(\omega+\Omega) \nonumber\\
&-& \tilde{T}_{62}(\omega)\tilde{T}_{26}(\omega+\Omega)-\tilde{T}_{82}(\omega)
\tilde{T}_{28}(\omega+\Omega)\Big],~~~~
\end{eqnarray}
\end{subequations}
in the regions A and B of BZ, respectively, with the T-matrix,
\begin{equation}
T^{(\alpha)}(\omega) = \left(
\begin{array}{ll}T_{AA}^{(\alpha)}(\omega)& T_{AB}^{(\alpha)}(\omega)\\
T_{BA}^{(\alpha)}(\omega)&T_{BB}^{(\alpha)}(\omega)\end{array}
\right),
\end{equation}
where the matrixes $T_{\mu\nu}^{(\alpha)}(\omega)$ ($\mu,\nu = A, B$) with the corresponding
matrix elements have been given explicitly in Ref. \onlinecite{Zeng22}. Moreover, a general
formalism is satisfied by $\tilde{T}_{jn}(\omega)\tilde{T}_{nj}(\omega+\Omega)$ as,
\begin{eqnarray}\label{T-Eq1}
\tilde{T}_{jn}(\omega)\tilde{T}_{nj}(\omega+\Omega)=
\sum\limits_{\alpha,\,\beta = 0}^{3}\tau_{\alpha}
T^{(\alpha)}_{jn}(\omega)\tau_{\beta}T^{(\beta)}_{nj}(\omega+\Omega)=
\sum\limits_{\alpha,\,\beta,\gamma = 0}^{3}i\bar{\epsilon}_{\alpha\beta\gamma}
T^{(\alpha)}_{jn}(\omega)T^{(\beta)}_{nj}(\omega+\Omega)\tau_{\gamma},
\end{eqnarray}
with $i\bar{\epsilon}_{\alpha\beta\gamma}$ that is defined as,
\begin{eqnarray}\label{eps-expression}
i\bar{\epsilon}_{\alpha\beta\gamma}=\delta_{\alpha\beta}\delta_{\gamma0}+(1-\delta_{\alpha0})
\delta_{\beta0}\delta_{\gamma\alpha}+\delta_{\alpha0}(1-\delta_{\beta0})\delta_{\gamma\beta}
+i\epsilon_{\alpha\beta\gamma},
\end{eqnarray}
where $\epsilon_{\alpha\beta\gamma}$ is the Levi-Civita tensor, and then
$i\bar{\epsilon}_{\alpha\beta\gamma}$ satisfies the following identities:
$\tau_{\alpha}\tau_{\beta}=\sum\limits_{\gamma}i\bar{\epsilon}_{\alpha\beta\gamma}\tau_{\gamma}$
and $i\bar{\epsilon}_{\alpha\beta\gamma}=i\bar{\epsilon}_{\gamma\alpha\beta}$. With the help of
the above general formalism (\ref{T-Eq1}), the relations in Eq. (\ref{To-sum-o}) can be derived
as,
\begin{subequations}\label{calculation-process}
\begin{eqnarray}
{n_{i}N\over\cos^{2}\theta_{\rm F}}\sum_{j''\in {\rm odd}}\hat{k}_{x}^{(1)}\cdot
\hat{k}_{\rm F}^{(j'')}\tilde{T}_{j''1}(\omega)\tilde{T}_{1j''}(\omega+\Omega) &=&
\sum_{\gamma}C^{(x)}_{A1}(\gamma)\tau_{\gamma},\\
C^{(x)}_{A1}(\gamma)=n_{i}N\sum\limits_{\substack{\alpha,\,\beta = 0}}^{3}
i\bar{\epsilon}_{\alpha\beta\gamma}\big[ T^{(\alpha)}_{11}(\omega)T^{(\beta)}_{11}(\omega+\Omega)
&+&T^{(\alpha)}_{31}(\omega)T^{(\beta)}_{13}(\omega+\Omega)-T^{(\alpha)}_{51}(\omega)
T^{(\beta)}_{15}(\omega+\Omega)\nonumber\\
&-&T^{(\alpha)}_{71}(\omega)T^{(\beta)}_{17}(\omega+\Omega)\big],~~~~\\
{n_{i}N\over\sin^{2}\theta_{\rm F}}\sum_{j''\in {\rm odd}}\hat{k}_{x}^{(2)}\cdot
\hat{k}_{\rm F}^{(j'')}\tilde{T}_{j''1}(\omega)\tilde{T}_{1j''}(\omega+\Omega) &=&
\sum_{\gamma}C^{(x)}_{A2}(\gamma)\tau_{\gamma},\\
C^{(x)}_{A2}(\gamma)=\cot\theta_{\rm F}n_{i}N\sum\limits_{\substack{\alpha,\,\beta = 0}}^{3}
i\bar{\epsilon}_{\alpha\beta\gamma}\big[ T^{(\alpha)}_{12}(\omega)T^{(\beta)}_{21}(\omega+\Omega)
&+& T^{(\alpha)}_{32}(\omega)T^{(\beta)}_{23}(\omega+\Omega)-T^{(\alpha)}_{52}(\omega)
T^{(\beta)}_{25}(\omega+\Omega)\nonumber\\
&-&T^{(\alpha)}_{72}(\omega)T^{(\beta)}_{27}(\omega+\Omega)\big],~~~\\
{n_{i}N\over\cos^{2}\theta_{\rm F}}\sum_{j''\in {\rm even}}\hat{k}_{x}^{(1)}\cdot
\hat{k}_{\rm F}^{(j'')}\tilde{T}_{j''1}(\omega)\tilde{T}_{1j''}(\omega+\Omega)
&=& \sum_{\gamma}C^{(x)}_{B1}(\gamma)\tau_{\gamma}, \\
C^{(x)}_{B1}(\gamma)=n_{i}N\tan\theta_{\rm F}\sum\limits_{\substack{\alpha,\,\beta = 0}}^{3}
i\bar{\epsilon}_{\alpha\beta\gamma}\big[ T^{(\alpha)}_{21}(\omega)T^{(\beta)}_{12}(\omega+\Omega)
&+& T^{(\alpha)}_{41}(\omega)T^{(\beta)}_{14}(\omega+\Omega)-T^{(\alpha)}_{61}(\omega)
T^{(\beta)}_{16}(\omega+\Omega)\nonumber\\
&-&T^{(\alpha)}_{81}(\omega)T^{(\beta)}_{18}(\omega+\Omega)\big],~~~~~~~\\
{n_{i}N\over\sin^{2}\theta_{\rm F}}\sum_{j''\in {\rm even}}\hat{k}_{x}^{(2)}\cdot
\hat{k}_{\rm F}^{(j'')}\tilde{T}_{j''2}(\omega)\tilde{T}_{2 j''}(\omega+\Omega)
&=& \sum\limits_{\gamma}C^{(x)}_{B2}(\gamma)\tau_{\gamma}, \\
C^{(x)}_{B2}(\gamma)=n_{i}N\sum\limits_{\substack{\alpha,\,\beta = 0}}^{3}
i\bar{\epsilon}_{\alpha\beta\gamma}\big[ T^{(\alpha)}_{22}(\omega)T^{(\beta)}_{22}(\omega+\Omega)
&+& T^{(\alpha)}_{42}(\omega)T^{(\beta)}_{24}(\omega+\Omega)-T^{(\alpha)}_{62}(\omega)
T^{(\beta)}_{26}(\omega+\Omega)\nonumber\\
&-&T^{(\alpha)}_{82}(\omega)T^{(\beta)}_{28}(\omega+\Omega)\big].~~~~
\end{eqnarray}
\end{subequations}

Substituting the above results in Eq. (\ref{calculation-process}) into Eq. (\ref{Itr-Lambda-o})
and Eq. (\ref{Itr-Lambda-e}), $Tr[\tau_{0}\tilde{\Lambda}^{(\text{A})}_{x}(\omega,\Omega)]$ and
$Tr[\tau_{0}\tilde{\Lambda}^{(\text{B})}_{x}(\omega,\Omega)]$ can be obtained explicitly as,
\begin{subequations}\label{Tr-Lambda}
\begin{eqnarray}
Tr[\tilde{\Lambda}^{(\text{A})}_{x}(\omega,\Omega)]=\sum\limits_{\beta = 0}^{3}\Big\{ Tr\big[\tau_{\beta}[\tau_{0}+\tilde{\Lambda}^{(\text{A})}_{x}(\omega,\Omega)]\big]
R_{\text{A}1\beta}^{(x)}(\omega,\omega+\Omega)+Tr\big[\tau_{\beta}[\tau_{0}
+\tilde{\Lambda}^{(\text{B})}_{x}(\omega,\Omega)]\big]
R_{\text{B}1\beta}^{(x)}(\omega,\omega+\Omega)\Big\},~~~~~~~\label{Tr-Lambda-A}\\
Tr[\tilde{\Lambda}^{(\text{B})}_{x}(\omega,\Omega)]=\sum\limits_{\beta = 0}^{3}\Big\{
Tr\big[\tau_{\beta}[\tau_{0}+\tilde{\Lambda}^{(\text{A})}_{x}(\omega,\Omega)]\big]
R_{\text{A}2\beta}^{(x)}(\omega,\omega+\Omega)+Tr\big[\tau_{\beta}[\tau_{0}
+\tilde{\Lambda}^{(\text{B})}_{x}(\omega,\Omega)]\big]
R_{\text{B}2\beta}^{(x)}(\omega,\omega+\Omega)\Big\},~~~~~~~\label{Tr-Lambda-B}
\end{eqnarray}
\end{subequations}
respectively, with the functions,
\begin{subequations}
\begin{eqnarray}
&&R_{\text{A}1\beta}^{(x)}(\omega,\omega+\Omega)=\sum\limits_{\gamma =0}^{3}
C^{(x)}_{\text{A}1}(\gamma)\tilde{I}_{\gamma}^{(\text{A})}(\beta,\omega,\omega+\Omega),~~~~~~~
R_{\text{A}2\beta}^{(x)}(\omega,\omega+\Omega)=\sum\limits_{\gamma =0}^{3}
C^{(x)}_{\text{A}2}(\gamma)\tilde{I}_{\gamma}^{(\text{A})}(\beta,\omega,\omega+\Omega),~~~~\\
&&R_{\text{B}1\beta}^{(x)}(\omega,\omega+\Omega)=\sum\limits_{\gamma =0}^{3}
C^{(x)}_{\text{B}1}(\gamma)\tilde{I}_{\gamma}^{(\text{B})}(\beta,\omega,\omega+\Omega),~~~~~~~
R_{\text{B}2\beta}^{(x)}(\omega,\omega+\Omega)=\sum\limits_{\gamma =0}^{3}
C^{(x)}_{\text{B}2}(\gamma)\tilde{I}_{\gamma}^{(\text{B})}(\beta,\omega,\omega+\Omega).~~~~
\end{eqnarray}
\end{subequations}

Now we turn to evaluate the similar traces of the product between the vertex kernel
$\tilde{\Lambda}^{(\text{A})}_{x}(\omega,\Omega)$ and matrix $\tau_{\alpha}$ with
$\alpha=1,2,3$ in the region A and the product of the vertex kernel
$\tilde{\Lambda}^{(\text{B})}_{x}(\omega,\Omega)$ and matrix $\tau_{\alpha}$ in the region B
in the kernel function (\ref{current-density-2}), where the derivation processes are almost
the same as the derivation processes for the above
$Tr[\tau_{0}\tilde{\Lambda}^{(\text{A})}_{x}(\omega,\Omega)]$ in Eq. (\ref{Tr-Lambda-A}) and
$Tr[\tau_{0}\tilde{\Lambda}^{(\text{B})}_{x}(\omega,\Omega)]$ in Eq. (\ref{Tr-Lambda-B}),
and the obtained results can be expressed explicitly as,
\begin{subequations}\label{Tr-tau-Lambda}
\begin{eqnarray}
Tr[\tau_{\alpha}\tilde{\Lambda}^{(\text{A})}_{x}(\omega,\Omega)]
&=&\sum\limits_{\substack{\beta = 0}}^{3}\Big\{
Tr\big[\tau_{\beta}[\tau_{0}+\tilde{\Lambda}^{(\text{A})}_{x}(\omega,\Omega)]\big]
R_{\text{A}1\beta}^{(x)}(\alpha,\omega,\omega+\Omega)
+ Tr\big[\tau_{\beta}[\tau_{0}+\tilde{\Lambda}^{(\text{B})}_{x}(\omega,\Omega)]\big]
R_{\text{B}1\beta}^{(x)}(\alpha,\omega,\omega+\Omega)\Big\},\nonumber\\
~~\\
Tr[\tau_{\alpha}\tilde{\Lambda}^{(\text{B})}_{x}(\omega,\Omega)]
&=&\sum\limits_{\substack{\beta = 0}}^{3}\Big\{
Tr\big[\tau_{\beta}[\tau_{0}+\tilde{\Lambda}^{(\text{A})}_{x}(\omega,\Omega)]\big]
R_{\text{A}2\beta}^{(x)}(\alpha,\omega,\omega+\Omega)
+ Tr\big[\tau_{\beta}[\tau_{0}+\tilde{\Lambda}^{(\text{B})}_{(x)}(\omega,\Omega)]\big]
R_{\text{B}2\beta}^{(x)}(\alpha,\omega,\omega+\Omega)\Big\},\nonumber\\
~~
\end{eqnarray}
\end{subequations}
with the functions,
\begin{subequations}
\begin{eqnarray}
R_{\text{A}1\beta}^{(x)}(\alpha,\omega,\omega+\Omega)&=&\sum\limits_{\lambda = 0}^{3}
C^{(x)}_{\text{A}1\alpha}(\lambda)\tilde{I}^{(\text{A})}_{\lambda}(\beta,\omega,\omega+\Omega),\\
C^{(x)}_{\text{A}1\alpha}(\lambda)&=&n_{i}N\sum\limits_{\substack{\mu,\,\nu = 0}}^{3}
\big(\sum\limits_{\sigma}i\bar{\epsilon}_{\mu\nu\sigma}i\bar{\epsilon}_{\sigma\alpha\lambda}\big)
\eta_{\alpha}(\nu)[T^{(\mu)}_{11}(\omega)T^{(\nu)}_{11}(\omega+\Omega)+T^{(\mu)}_{31}(\omega)
T^{(\nu)}_{13}(\omega+\Omega)\nonumber\\
&-& T^{(\mu)}_{51}(\omega)T^{(\nu)}_{15}(\omega+\Omega)-T^{(\mu)}_{71}(\omega)
T^{(\nu)}_{17}(\omega+\Omega)],\\
R_{\text{B}1\beta}^{(x)}(\alpha,\omega,\omega+\Omega)&=&\sum\limits_{\lambda = 0}^{3}
C^{(x)}_{\text{B}1\alpha}(\lambda)\tilde{I}^{(\text{B})}_{\lambda}(\beta,\omega,\omega+\Omega),\\
C^{(x)}_{\text{B}1\alpha}(\lambda) &=& n_{i}N\tan\theta_{\rm F}
\sum\limits_{\substack{\mu,\,\nu = 0}}^{3}\big(\sum\limits_{\sigma}i\bar{\epsilon}_{\mu\nu\sigma}
i\bar{\epsilon}_{\sigma\alpha\lambda}\big)\eta_{\alpha}(\nu)\big[T^{(\mu)}_{21}(\omega)
T^{(\nu)}_{12}(\omega+\Omega)+T^{(\mu)}_{41}(\omega)T^{(\nu)}_{14}(\omega+\Omega)\nonumber\\
&-& T^{(\mu)}_{61}(\omega)T^{(\nu)}_{16}(\omega+\Omega)-T^{(\mu)}_{81}(\omega)
T^{(\nu)}_{18}(\omega+\Omega)\big], \\
R_{\text{A}2\beta}^{(x)}(\alpha,\omega,\omega+\Omega)&=&\sum\limits_{\lambda = 0}^{3}
C^{(x)}_{\text{A}2\alpha}(\lambda)\tilde{I}^{(\text{A})}_{\lambda}(\beta,\omega,\omega+\Omega),\\
C^{(x)}_{\text{A}2\alpha}(\lambda) &=& n_{i}N\cot\theta_{\rm F}
\sum\limits_{\substack{\mu,\,\nu = 0}}^{3}\big(\sum\limits_{\sigma}i\bar{\epsilon}_{\mu\nu\sigma}
i\bar{\epsilon}_{\sigma\alpha\lambda}\big)\eta_{\alpha}(\nu)\big[T^{(\mu)}_{12}(\omega)
T^{(\nu)}_{21}(\omega+\Omega)+T^{(\mu)}_{32}(\omega)T^{(\nu)}_{23}(\omega+\Omega)\nonumber\\
&-& T^{(\mu)}_{52}(\omega)T^{(\nu)}_{25}(\omega+\Omega)-T^{(\mu)}_{72}(\omega)
T^{(\nu)}_{27}(\omega+\Omega)\big], \\
R_{\text{B}2\beta}^{(x)}(\alpha,\omega,\omega+\Omega)&=&\sum\limits_{\lambda = 0}^{3}
C^{(x)}_{\text{B}2\alpha}(\lambda)\tilde{I}^{(\text{B})}_{\lambda}(\beta,\omega,\omega+\Omega),
\end{eqnarray}
\begin{eqnarray}
C^{(x)}_{\text{B}2\alpha}(\lambda) &=& n_{i}N\sum\limits_{\substack{\mu,\,\nu = 0}}^{3}
\big(\sum\limits_{\sigma}i\bar{\epsilon}_{\mu\nu\sigma}
i\bar{\epsilon}_{\sigma\alpha\lambda}\big)\eta_{\alpha}(\nu)\big[T^{(\mu)}_{22}(\omega)
T^{(\nu)}_{22}(\omega+\Omega)+T^{(\mu)}_{42}(\omega)T^{(\nu)}_{24}(\omega+\Omega)\nonumber\\
&-& T^{(\mu)}_{62}(\omega)T^{(\nu)}_{26}(\omega+\Omega)-T^{(\mu)}_{82}(\omega)
T^{(\nu)}_{28}(\omega+\Omega)\big],
\end{eqnarray}
\end{subequations}
where $\sum\limits_{\sigma}i\bar{\epsilon}_{\mu\nu\sigma}i\bar{\epsilon}_{\sigma\alpha\lambda}$
satisfies the following identity,
\begin{eqnarray}
\sum\limits_{\sigma}i\bar{\epsilon}_{\mu\nu\sigma} i\bar{\epsilon}_{\sigma\alpha\lambda}
&=& -4\delta_{\mu0}\delta_{\nu0}\delta_{\alpha0}\delta_{\lambda0}
+\delta_{\alpha\mu}\delta_{\lambda0}\delta_{\nu0}+\delta_{\alpha\nu}\delta_{\mu0}\delta_{\lambda0}
+\delta_{\lambda\mu}\delta_{\nu0}\delta_{\alpha0}+\delta_{\mu0}\delta_{\alpha0}\delta_{\lambda\nu}
+\delta_{\alpha\lambda}\delta_{\mu\nu}+\delta_{\alpha\nu}\delta_{\lambda\mu}
-\delta_{\alpha\mu}\delta_{\lambda\nu} \nonumber\\
&+& i\delta_{\alpha0}\epsilon_{\lambda\mu\nu}+i\delta_{\lambda0}\epsilon_{\alpha\mu\nu}
+ i\delta_{\mu0}\epsilon_{\nu\alpha\lambda}+i\delta_{\nu0}\epsilon_{\mu\alpha\lambda},
\end{eqnarray}
and the tensor $\eta_{\alpha}(\nu)$ is defined as,
\begin{equation}
\eta_{\alpha}(\nu) = \left\{
\begin{array}{ll}
1,&\nu = 0, \alpha \\
-1, &{\rm others} \;.
\end{array}
\right.
\end{equation}
Substituting the above results in Eqs. (\ref{Tr-Lambda}) and (\ref{Tr-tau-Lambda}) into
Eq. (\ref{current-density-2}) of the main text, we therefore obtain the kernel function
$J_{xx}(\omega,\omega+\Omega)$ in Eq. (\ref{current-density-2}) of the main text.


\end{appendix}

\end{widetext}


\end{document}